%% file: main.tex
\documentclass[preprint2]{aastex631}
\usepackage{gensymb}
\usepackage{comment}
\usepackage[utf8]{inputenc}
\usepackage{url}
\usepackage{amsmath}
\usepackage{epstopdf}
\usepackage{wrapfig}
\usepackage{xcolor}
\usepackage[normalem]{ulem}
\usepackage{soul}

\begin{document}

\title{The Near Infrared Imager and Slitless Spectrograph for the  {\it James Webb Space Telescope} - III. Single Object Slitless Spectroscopy}

\correspondingauthor{Lo\"ic Albert}
\email{loic.albert@umontreal.ca}

\input{authors.tex}
\begin{abstract}
The Near Infrared Imager and Slitless Spectrograph instrument (NIRISS) is the Canadian Space Agency (CSA) contribution to the suite of four science instruments of JWST. As one of the three NIRISS observing modes, the Single Object Slitless Spectroscopy (SOSS) mode is tailor-made to undertake time-series observations of exoplanets to perform transit spectroscopy. The SOSS permits observing point sources between 0.6 and 2.8\,$\mu$m at a resolving power of 650 at 1.25~$\mu$m using a slit-less cross-dispersing grism while its defocussing cylindrical lens enables observing targets as bright as $J$=6.7 by spreading light across 23 pixels along the cross-dispersion axis. This paper officially presents the design of the SOSS mode, its operation, characterization, and its performance, from ground-based testing and flight-based Commissioning. On-sky measurements demonstrate a peak photon conversion efficiency of 55\% at 1.2\,$\mu$m. The first time-series on the A-type star BD$+$60$^\circ$1753 achieves a flux stability close to the photon-noise limit, so far tested to a level of 20 parts per million on 40-minute time-scales after simply subtracting a long-term trend. Uncorrected 1/f noise residuals underneath the spectral traces add an extra source of noise equivalent to doubling the readout noise. Preliminary analysis of a HAT-P-14b transit time-series indicates that it is difficult to remove all the noise in pixels with partially saturated ramps. Overall, the SOSS delivers performance at the level required to tackle key exoplanet science programs such as detecting secondary atmospheres on terrestrial planets and measuring abundances of several chemical species in gas giants.

\textit{Subject headings: } instrumentation: spectrographs – methods: data analysis – techniques: spectroscopic
\end{abstract}

\section{Introduction} \label{sec:intro}
Time-series observations (TSOs) are a powerful method to study the atmosphere of exoplanets or map their surface. Transmission spectroscopy as a means to study the chemical composition of exoplanet atmospheres was proposed by \cite{seager.2000} soon after observations of the first exoplanet transit of HD~209458b \citep{charbonneau.2000, henry.1999, henry.2000}. Detection of additional exoplanets through their transit \citep{konacki.2003,bouchy.2004,alonso.2004} paved the way to the ground-based transit surveys \citep{bakos.2004,pollacco.2006} and to space missions like CoRoT \citep{auvergne.2009} and Kepler \citep{borucki.2010} which is responsible for the majority of today's known exoplanets. As a discovery method, transits are restricted to a single photometric band-pass, but they open access to a key planet parameter: radius. When complemented with radial velocity masses, planet densities could be estimated for the first time. This confirmed, for example, that gas giants followed a linear density versus mass power law \citep{hatzes.2014}. By multiplexing the transit observation, transmission spectroscopy would soon allow studying the change of apparent radius as a function of wavelength brought about by molecule opacities in the exoplanet atmospheres. The first attempt at transmission spectroscopy from the ground successfully identified sodium in the atmosphere of HD~189733\,b \citep{redfield.2008}. However, challenges associated with scintillation meant that achieving the required measurement accuracies of less than $\sim$100 ppm with ground-based observations had the community turn to the space-borne Hubble Space Telescope (HST). The first attempt at transit spectra with HST Space Telescope Imaging Spectrograph (STIS) observations \citep{pont.2007} was followed by haze detection in HD~209458\,b \citep{pont.2008}, water vapor in HD~189733\,b \citep{swain.2008}. Once the sample of hot Jupiters probed with WFC3 became large enough, not only was the presence of water vapor found to be ubiquitous but the depth of its absorption band to be modulated by the presence of high altitude clouds \citep{sing.2016}. A few more molecules and atoms were detected with transit or secondary eclipse spectroscopy. Notably, using NICMOS on HST, \cite{swain.2009} found the first evidence of carbon dioxide near 2\,$\mu$m in the atmosphere of HD~189733b. At longer wavelengths (4.5, 8.0 and 24~$\mu$m), Spitzer led to the first albedo and hot-side temperature measurements of hot jupiters through secondary eclipse TSOs \citep{deming.2005,charbonneau.2005}, enabling study of heat transport in these planets. Ultimately, Spitzer and HST's limited mirror sizes or infrared wavelength coverage could not realistically detect the atmosphere of small, cold terrestrial planets which the TESS survey \citep{ricker.2015} was about to discover around bright nearby stars.


The Single Object Slitless Spectroscopy (SOSS) observing mode was conceived as a response to the constraints of exoplanet transmission spectroscopy. It emerged as a pivotal science mode for the rescope of the Tunable Filter Imager \citep{doyon.2012} into the Near Infrared Imager and Slitless Spectrograph (NIRISS) \citep{doyon.2023}, in 2011. By then, the instrumentation suite of the James Webb Space Telescope (JWST) -- which had been designed prior to the rise of transmission spectroscopy -- was already in its integration and testing phase. The SOSS was designed to enable NIRISS observations of TESS exoplanets, whose hosts would be several magnitudes brighter than the mean exoplanets hosts known from ground-based surveys and Kepler. Inspired by HST's scanning mode, something that JWST can not reproduce -- non-sidereal scanning speed is limited to $\leq 1$~pixel/second -- we introduced a defocussing lens in SOSS. Its cylindrical lens pushes detector saturation to $\sim2-3$ mag brighter targets by spreading flux over 23 pixels along the spatial axis of the spectroscopic trace. Also, the spectral coverage of SOSS ($0.6\leq \lambda \leq 2.8~\mu$m) was chosen to bracket the peak spectral energy distribution (SED) of most planetary host stars where maximum signal-to-noise ratio (SNR) is expected. It encompasses several water, methane, CO, and CO$_2$ bands as well as potassium and helium atomic lines. The SOSS mode offers complementary wavelength coverage in the blue near-infrared ($\lambda \leq 2.8\mu$m) to what the Near Infrared Spectrograph (NIRSpec) grating \citep{birkmann.2022} offers redward of $3.0\mu$m, at similar resolving power. 

This paper is part of a series that present NIRISS and its observing modes: the NIRISS overview paper \citep{doyon.2023}; Wide-Field Slitless Spectroscopy \citep{willott.2022}, Aperture Masking Interferometry \citep{sivaramakrishnan.2023} and Kernel Phase interferometry \citep{kammerer.2022}. Some of the characterization of the SOSS mode, particularly that of the detector, was obtained during the third cryogenic vaccum ground testing campaign (CV3) conducted at the Goddard Space Flight Center in 2016, the rest is from instrument in-flight Commissioning in 2022.

This SOSS mode paper starts with overviews of the hardware (Sec.~\ref{sec:hardware}), the optical design (Sec.~\ref{sec:opticaldesign}), and the detector (Sec.~\ref{sec:detector_overview}). Performance from ground-based observations are described in Sec.~\ref{sec:groundperformance} while
the operations' concept is detailed in Sec.~\ref{sec:opsconcept}. 
Detector characterization is inserted in Sec.~\ref{sec:detector_characterization}
before performance from Commissioning observations (Sec.~\ref{sec:flight}). 
The instrument team simulator is presented for the first time in Sec.~\ref{sec:idtsoss}, followed by the time-series noise performance (Sec.~\ref{sec:noiseperformance}). We finally summarize the SOSS mode salient points in Sec.~\ref{sec:conclusions} and quickly present the \emph{NIRISS Exploration of the Atmospheric diversity of Transiting exoplanets} (NEAT) Guaranteed Time Observations program.

\section{Hardware Overview} \label{sec:hardware}
NIRISS is the result of a late design change due to the difficultly implementing the Fabry-Perot interference plates at the core of the Tunable Filter Imager (TFI) -- the name of the instrument prior to its rescope in 2011 -- See \cite{haley.2012} , \cite{doyon.2023}. The Fabry-Perot was removed and most of the optical components supporting its function in two wheels, the pupil wheel and the filter wheel, in the instrument's collimated beam were exchanged for other optical components.  Nine circular slots are available for each wheel. The NIRISS optics provides an image of the JWST primary mirror (pupil image) with a clear aperture  equivalent to 39 mm in diameter at the location of the pupil wheel. Each wheel is rotated by a stepper motor with a single-step resolving accuracy of 0\fdg1651 at the pupil wheel and 0\fdg1585 at the filter wheel.
 
NIRISS uses a Three-Mirror Assembly for the collimator and another Three-Mirror Assembly for the imaging optics. A Pick Off Mirror of rectangular shape selects a field of view slightly oversized with respect to the detector \ref{fig:layout}). Four occulting spots of various diameters, relics of TFI, are engraved in the Pick Off Mirror and produce dark spots on the detector in Imaging mode near the footprint of the SOSS traces (See their description in \cite{willott.2022}). Fortunately, these spots are not a concern for SOSS as they fall off the field of view when dispersed by the SOSS grism (whose description follows in Sec.~\ref{sec:opticaldesign}). Special care is taken to account for these in the flat field and non-linearity calibration images, patching them with dispersed images obtained with a low-resolution grism (GR150). Broadband sources of illumination, relics of the TFI design, produce highly non-uniform illumination on the detector and their stability is poor so their usefulness is very limited.

\section{Optical Design and Detector Overview} \label{sec:opticaldesign}
\begin{figure}[h]
    \centering
    \includegraphics[width=\linewidth]{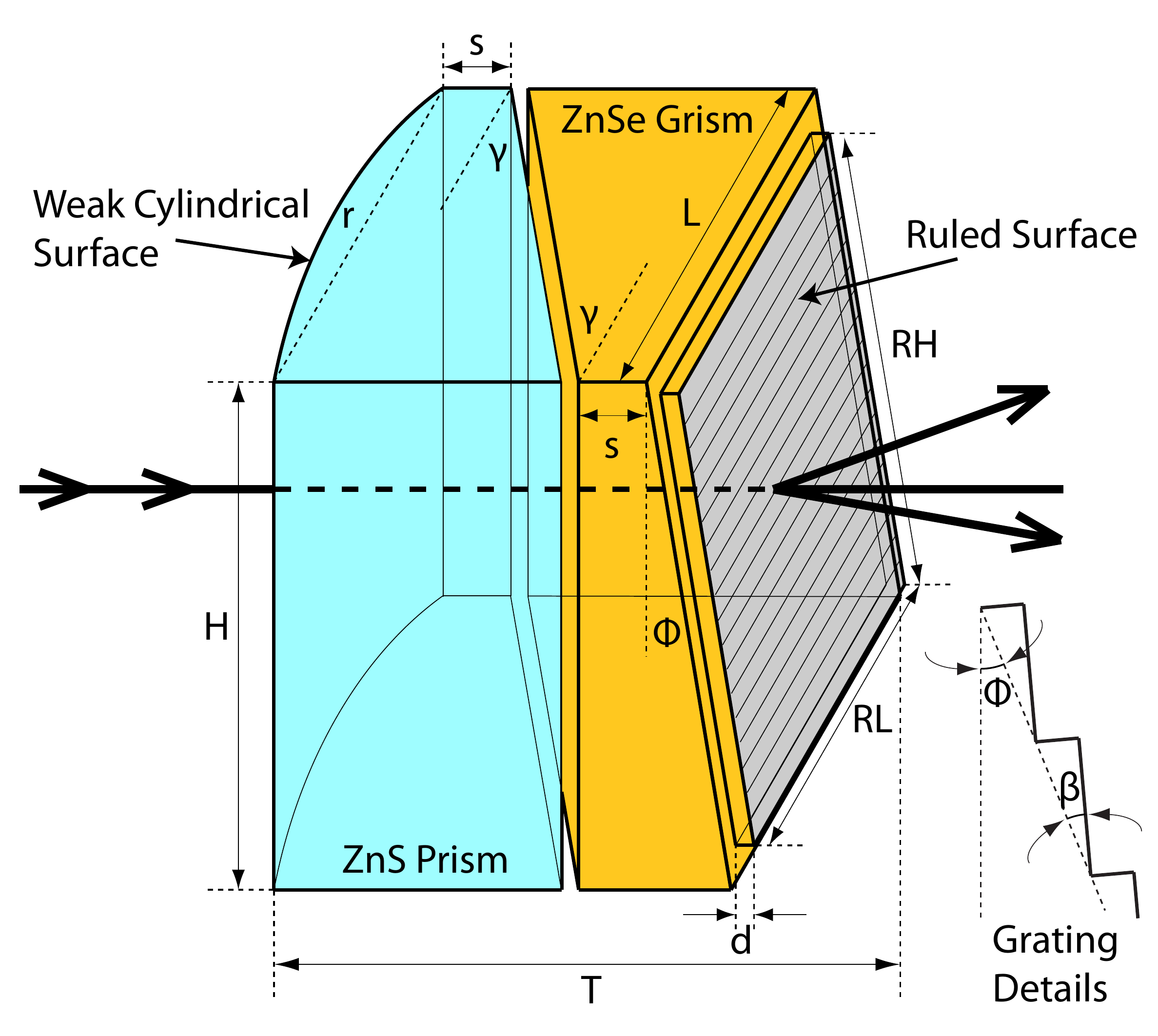}
    \caption{Schematic (not to scale) of the GR700XD. It consists of a zinc selenide (ZnSe) grism, a zinc sulfide (ZnS) prism that serves as a cross-disperser to split orders apart, and a weak cylindrical lens on the entrance surfaceof the ZnS prism designed to spread the PSF along the cross-dispersion axis. As-built specifications are given in Tab.~\ref{tab:gr700xd}.}
    \label{fig:GR700XD}
\end{figure}

The SOSS mode was designed to observe single point-source targets. {\em De facto}, it had to be slitless because no re-imaging optics necessary to place a slit existed in the instrument when the SOSS mode was developed. Science observations use the GR700XD element, a grating prism (GR) yielding a spectral resolution of roughly 700 with cross-dispersion (XD) of its spectral orders (See Fig.~\ref{fig:GR700XD}) It is inserted in the Pupil Wheel along with the CLEAR element (an open position with an oversized circular mask) in the Filter Wheel. The GR700XD element itself has three components: 1) the zinc selenide (ZnSe) grism onto which a grating was machined at the Lawrence Livermore National Laboratory \citep{kuzmenko.2014}; 2) the zinc sulfide (ZnS) prism used to cross-disperse orders in the spatial direction, orthogonal to the dispersion direction; and 3) a square mask to ensure that all through paths are diffracted by the grating. We selected the ZnS and ZnSe infrared glasses primarily for their high refractive indices, allowing larger spectral dispersion and resolving power, but also because they maximized the ability to spread orders apart along the cross-dispersion axis.The ZnS prism was manufactured by BMV Optical Technologies (Ottawa ON) and coated by Thin Film Labs (Milford PA) while the ZnSe grism was coated by II-VI Optical Systems on both surfaces.

Other important features were included in the design. First, the front surface of the ZnS prism has a convex cylindrical shape with a radius of curvature of 25.3\,m (7$\lambda$ over 30\,mm at $\lambda=632$~nm) to spread the beam by roughly 23\,pixels along the spatial direction. This is to prevent saturation when observing bright stars and to reduce flat-field errors. Second, we dialed a small rotation of the mask, prism, and grism into its mechanical cell about the optical axis in order to introduce a small 3$\degree$ tilt of monochromatic light with respect to the detector columns. The goal was to benefit from the 23-pixel trace extent along the spatial axis to enable spectral resampling, resulting in a resolving power improvement. However, later, in order to properly fit the whole first-order trace within the smallest science subarray, fine-tuning of the Pupil Wheel position at ground testing resulted in a rotation in the opposite direction that nearly cancels the designed monochromatic tilt. In the end, the apparent tilt varies slightly with wavelength ($\sim0^\circ \leq {\rm tilt} \leq 2^\circ$).

\begin{table}
    \centering
    \begin{tabular}{p{0.4\linewidth}p{0.2\linewidth}p{0.1\linewidth}p{0.2\linewidth}}
    \tableline
    \tableline
    & Grism & & Prism  \\
    \tableline
    
    Material & ZnSe & 									& ZnS \\
    Prism apex angle ($\phi$) & $1.9\pm0.1^\circ$ &			& $1.9\pm0.1^\circ$ \\
    Prism length (L) & 33.0\,mm & 							&  33.0\,mm \\
    Prism height (H) & 33.0\,mm &							& 33.0\,mm \\
    Thin side (s) & 3.0\,mm &								& 3.07\,mm \\
    Grating density ($\rho$) & 54.3\,l/mm (at 300\,K) &  			&  \\ 
    Facet blaze angle ($\beta$) & $2.63\pm0.2^\circ$ & 			&  \\
    Ruled length (RL) & 28.9\,mm & 						&  \\
    Ruled height (RH) & 27.9\,mm &						&  \\
    Recessed thickness (d) & 1.0\,mm &						&  \\
    Cylinder radius of curv. (r) & &						& 25.30\,m \\
    Prism wedge ($\gamma$)& $10^\circ 26$\arcmin &					& $10^\circ 26$\arcmin \\
    Total thickness (T) & \multicolumn{3}{c}{13.0\,mm} \\
    Full pupil diameter & \multicolumn{3}{c}{39.0\,mm} \\
    \tableline
    \tableline
    \end{tabular}
    \caption{As-built GR700XD optics specifications.}
    \label{tab:gr700xd}
\end{table}

\begin{figure*}[h]
    \centering
    \includegraphics[width=\linewidth]{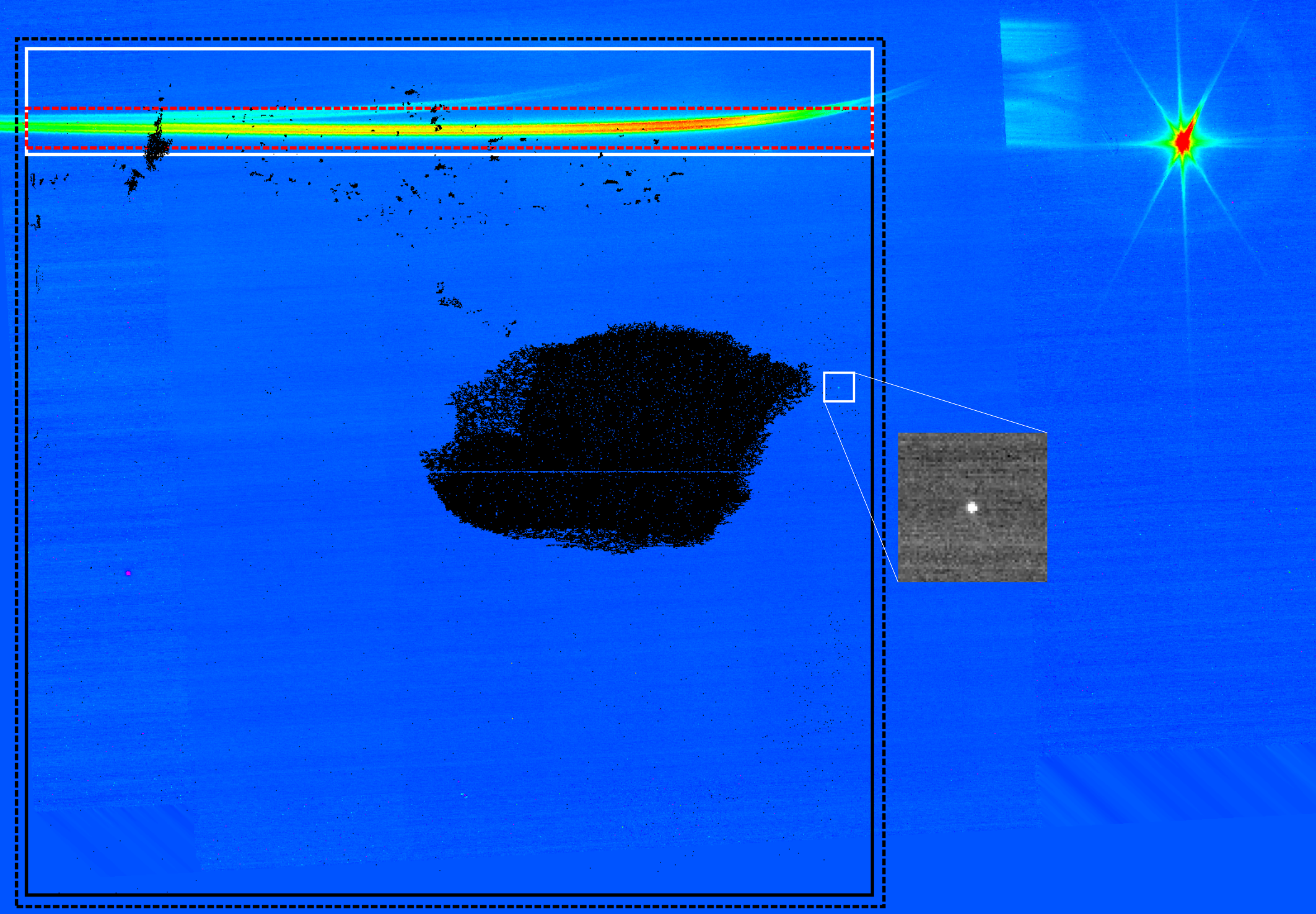}
     \caption{Image composite assembled from ground-based testing showing the detector footprint (solid black square) relative to the spectral diffraction orders (colored). For a target positioned at the nominal acquisition spot position (thumbnail), spectral orders 1, 2, and 3 lands on the detector while order 0 falls outside. Note that order 3 is too faint to be visible here. Three detector readouts are available: FULL, SUBSTRIP256 (white rectangle) and SUBSTRIP96 (dashed red rectangle, offset by 10 pixels relative to SUBSTRIP256). Overlaid in black is a map of the epoxy voids present in the detector. Besides the main void near the center, several small epoxy voids are present in the vicinity of the SOSS traces within the two science subarrays. In this image, the slow pixel readout axis is horizontal and the fast axis is vertical. The approximate field of view produced by the pick-off mirror is shown as a dashed black square. It is oversized by roughly 140 pixels all around the detector.}
    \label{fig:layout}
\end{figure*}

The GR700XD spreads three spectral orders\footnote{An extremely faint fourth order trace appears on deep stacks} onto the detector (See Fig.~\ref{fig:layout}) when the target is nominally positioned on the acquisition spot. The zeroth order falls outside the detector footprint. However,
orders 0,1 and 2 of field stars can also leave imprints on the detector in certain field positions, provided that the field star falls within the pick-off mirror field of view in direct imaging (extending roughly 140 pixels all around the detector edges). 

\begin{figure}
    \centering
    \includegraphics[width=\linewidth]{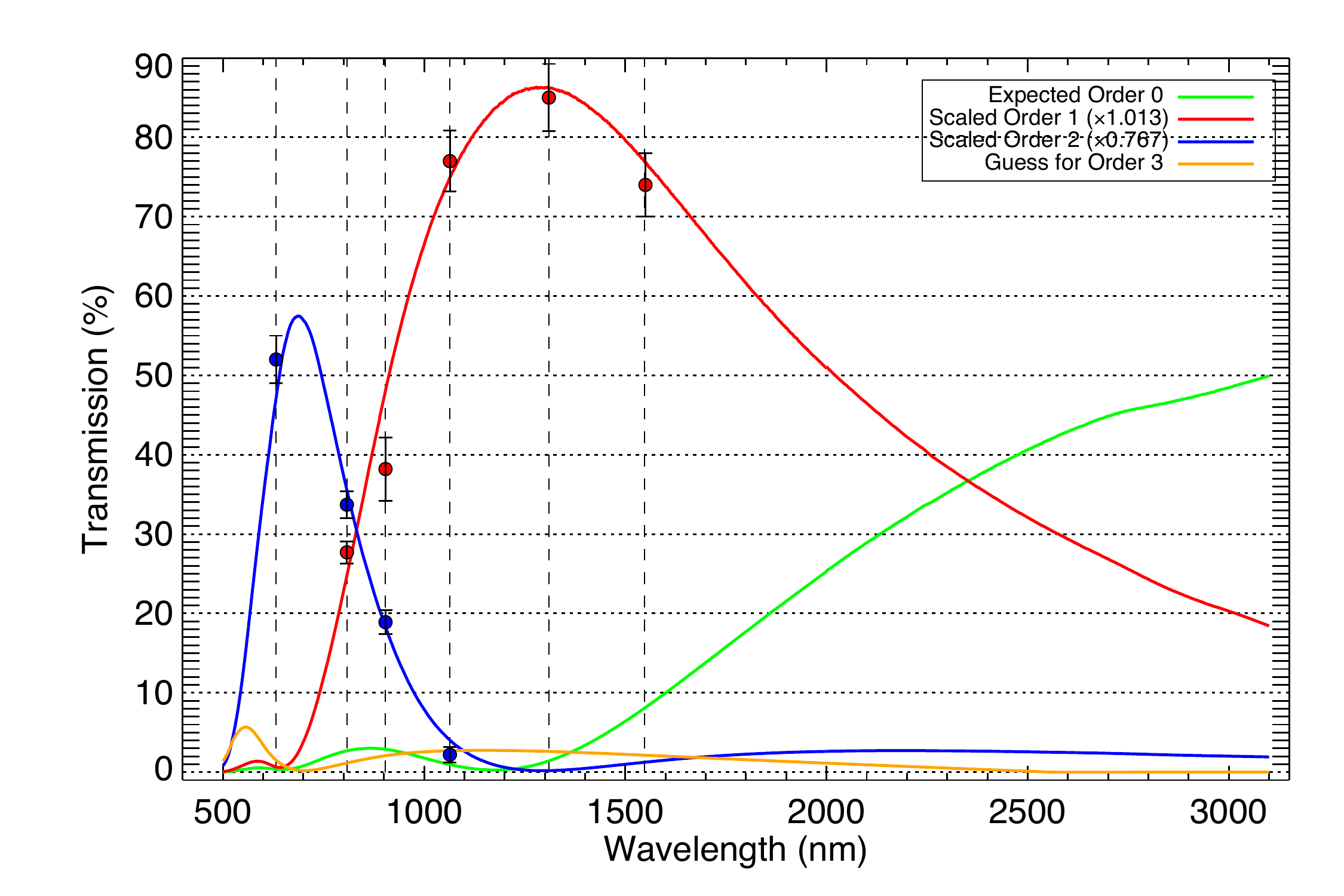}
    \caption{Transmission of the GR700XD optics measured in the lab (data points with error bars) compared with a PC Grate model (solid colored lines) calculated from first principles using the measured groove shape. The measurement was performed on the GR700XD with a circular 25 mm collimated beam so this measurement does not account for the losses due to the square mask, the secondary mirror, and the struts. In-flight transmission is expected to be lower by a factor of 0.663.}
    \label{fig:gr700xd}
\end{figure}

Most of the science content is expected to come from the first order ($0.84\,\mu{\rm m} \leq \lambda \leq 2.83\,\mu$m). The blaze wavelength of the grism is $\lambda_{b} = 1.23\,\mu$m with typical 40\% cutoffs from peak transmission at $\lambda_{-}=\frac{2}{3} \lambda_{b} \approx 0.82\,\mu$m and $\lambda_{+}=2\lambda_{b} \approx 2.46\,\mu$m. In the second order, the peak is expected near $\frac{1}{2} \lambda_{b} \approx 0.62\,\mu$m with 40\% cutoffs at $\lambda_{-}=\frac{4}{5} \lambda_{b} \approx 0.49\,\mu$m and $\lambda_{+}=\frac{4}{3}\lambda_{b} \approx 0.82\,\mu$m. Since the gold-coated mirrors cut off at $\approx0.6\,\mu$m, the effective second-order operable range is $\sim0.6\,\mu{\rm m} \leq \lambda \leq 0.82\,\mu$m, nicely complementing the first order in the blue. The throughput was maximized by applying anti-reflection coatings on all four grism and prism surfaces, yielding an end-to-end  measured GR700XD peak transmission of $\sim85\%$ at $1.3\,\mu$m in order 1 and $\sim55\%$ at $0.7\,\mu$m in order 2 (See Fig~\ref{fig:gr700xd}). Combined with a spectral dispersion roughly two times larger in order 2 ($\sim0.47$~nm/pixel) than in order 1 ($\sim0.97$~nm/pixel), the SNR per pixel in order 2 is expected to be about half that in order 1 near the peak of transmission. The third order is not intended to be of any scientific use, with the model transmission of only a few percent redward of 0.6\,$\mu$m.

A challenge associated with SOSS is the trace overlap between orders 1 and 2. This occurs because limited mechanical clearance (a maximum thickness of 13 mm) prevented the use of higher opening angles, $\phi$, for the prism and grism. Contamination, therefore, occurs at the red end of both traces, $\lambda \geq 2.2\,\mu$m in the first order and $\lambda \geq 1.1\,\mu$m in the second order. Typically, for a G star, intensity in the second order is 2--7\% that of the first order, per pixel, where the traces overlap. Clean regions of the spectrum (devoid of overlap for a box size of 30 pixels) roughly span $\sim0.84\,\mu{\rm m} \leq \lambda \leq 2.2\,\mu$m in order 1 and $\sim0.6\,\mu{\rm m} \leq \lambda \leq 1.1\,\mu$m in order 2. The \emph{trace overlap} contamination problem should not be mistaken for the \emph{field} contamination problem. Indeed, no slit is present (NIRISS has no image plane access) which means field contamination by other sources needs to be taken into account when planning observations. Such a tool exists:\url{https://exoctk.stsci.edu/contam_visibility}.

An engineering GR700XD grism built by Bach Corporation used during ground testing had the spectral traces oriented parallel to the fast readout axis to shorten frame time by using the 4-amplifier readout. However, an important optical ghost then appeared near the order 1 trace. The flight GR700XD built at Lawrence Livermore National Labs is therefore oriented perpendicular to the fast readout direction to prevent serious ghosts, but it also prevents use of a 4-amp readout multiplexing which would have shortened the frame time.

\begin{figure}
\centering
\includegraphics[width=\linewidth]{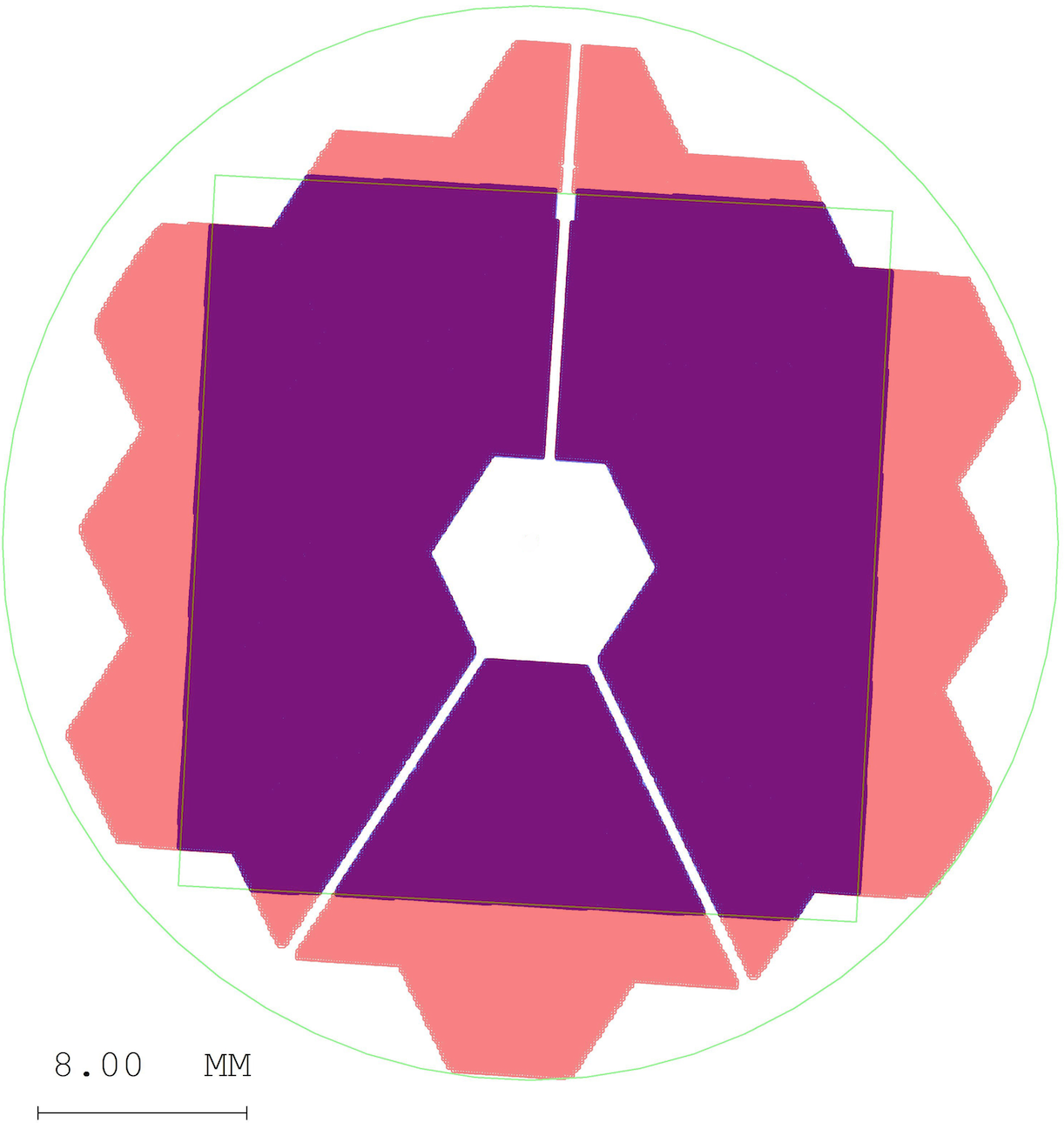}
\caption{\label{fig:pupilmask} The GR700XD pupil mask (in purple) relative to the NIRISS CLEAR mask pupil (pink) as modeled in \emph{CodeV} (post CV3 revision). The ratio of the mask to that of the open pupil is 0.663. The GR700XD elements and mask are rotated $3.0^\circ$ with respect to its holder and to the detector. The height of the square mask (green square) is 27.4~mm and its width is 25.8\,mm. For reference, the green circle is 40.45\,mm in diameter.}

\end{figure}

Due to mechanical clearance (GR700XD maximum total thickness of 13.0 mm) and manufacturing constraints (grating machining limited to square surfaces), the ruled surface of the grism is a square (26.0$\times$26.8\,mm), smaller than the full instrument pupil ($\sim39$~mm in diameter). The field mask at the entrance of the GR700XD ensures that light only reaches the ruled surface with a margin of about 1.0~mm around the edges. An important fraction of the full pupil light is therefore lost with the mask, secondary mirror, and strut obscurations resulting in a throughput fraction of 0.663. Fig.~\ref{fig:pupilmask} shows the final pupil of the GR700XD compared to that of the telescope. 

The ZnSe grism transmission was measured in the Universit\'e de Montr\'eal laboratory using LED sources. Our measurement setup produced a collimated beam of light in which we inserted the GR700XD element, both the ZnSe grism and the ZnS prism. We used a 25 mm circular stop at the entrance of the GR700XD, smaller than the 39 mm NIRISS pupil. The flux in the point spread function (PSF) was measured with and without the GR700XD in the beam on multiple trials. Because the test beam was smaller than the GR700XD square mask and because the secondary mirror and strut obscuration were not present in the lab, the transmission presented in Fig.~\ref{fig:gr700xd} should be multiplied by 0.663 to represent the in-flight performance expectations; roughly 56\% at the peak of order 1 and 36\% at the peak of order 2.

\section{Detector Overview} \label{sec:detector_overview}
The detector is a Teledyne H2RG detector with a 5$\,\mu$m cut-off controlled using the SIDECAR ASIC. Its performance and characteristics are detailed in \cite{doyon.2023}. It was tuned for a gain of $1.61\pm0.03$\,e$^-$/ADU (Analog-to-digital units). Its full well depth ($\sim$100\,000\,e$^-$) is sampled to no more than $\sim$80\,000\,e$^-$ based on a bias level in the 12-15 kADU range, and the onset of $\geq10$\% non-linearity or digital saturation occurs between 62-65 kADU. The readout noise (on a single frame) is 10.5 ADU ($\sim$17\,e$^-$). The detector harbors a large epoxy void near its center and smaller void regions towards the top where the SOSS traces are located (seen in black in Fig~\ref{fig:layout}). Other than having a slightly lower gain (by about 0.03\,e$^-$/ADU), pixels in epoxy voids appear to behave like other pixels. The detector non-linearity was characterized at CV3. Two lamps onboard NIRISS, legacy of the TFI instrument design \citep{doyon.2008}, do not have the flux stability to enable further updates of the flat field and non-linearity calibrations.

\section{Ground Based Optics Characterization}
\label{sec:groundperformance}

The ground-based NIRISS performance verification was carried out during CV3 in December 2015/January 2016 at Goddard Space Flight Center when the Instrument module was cooled to 40\,K. The JWST Optical Telescope Elements Optical Simulator provided point sources as well as uniform illumination capabilities, thus allowing an assessment of the SOSS optics performance.

\subsection{Spectral Trace Position}
\begin{figure*}
    \centering
    \includegraphics[width=\linewidth]{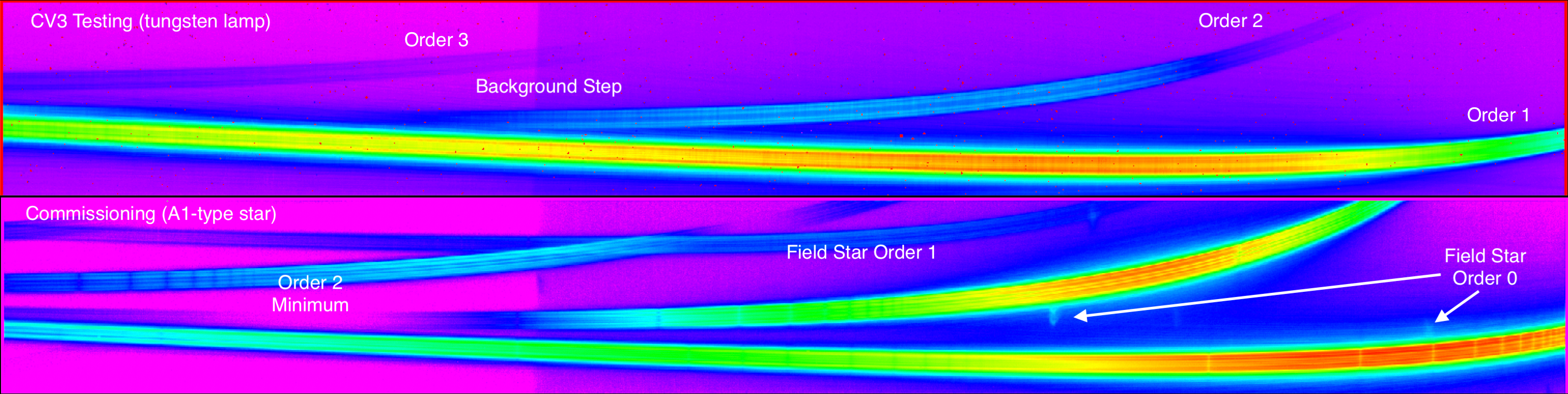}
    \caption{SOSS mode. First light at CV3 testing on a tungsten source at 1500\,K (top) and during Commissioning on the A-type flux standard BD$+$60$^\circ$1753 (bottom). Three spectral orders are produced by the GR700XD and they harbor a double-horn shape along their cross-dispersion (vertical) axis due to the cylindrical lens. Other notable features are: 1) the zodiacal background break near column 700 due to the background zeroth order falling off the Pick Off Mirror leftward; 2) the field star zeroth order and first order contamination; 3) the transmission minimum in order 2 which dampens the concern of inter-order contamination at their immediate left (red-ward).}
    \label{fig:firstlight}
\end{figure*}

First light with a 1500\,K tungsten source confirmed the presence of the bright diffracted order 1 accompanied by a slightly fainter order 2; both featuring a ``double-horn'', defocused, trace profile along the cross-dispersion (spatial) axis as expected (See Fig.~\ref{fig:firstlight}).
Order 3 was detected but is faint. The measured trace position and curvature were within approximately two pixels of those expected from the {\em Code V} optics model.

The overall trace orientation was fine-tuned by rotating the filter wheel a few degrees away from the nominal position in order to capture as much of the first spectral diffraction order as possible within the smallest subarray: 96 pixels wide. We ensured that the final chosen orientation, $+1.1\degree$, did not produce pupil vignetting or transmission loss.

\begin{figure}
    \centering
    \includegraphics[width=\linewidth]{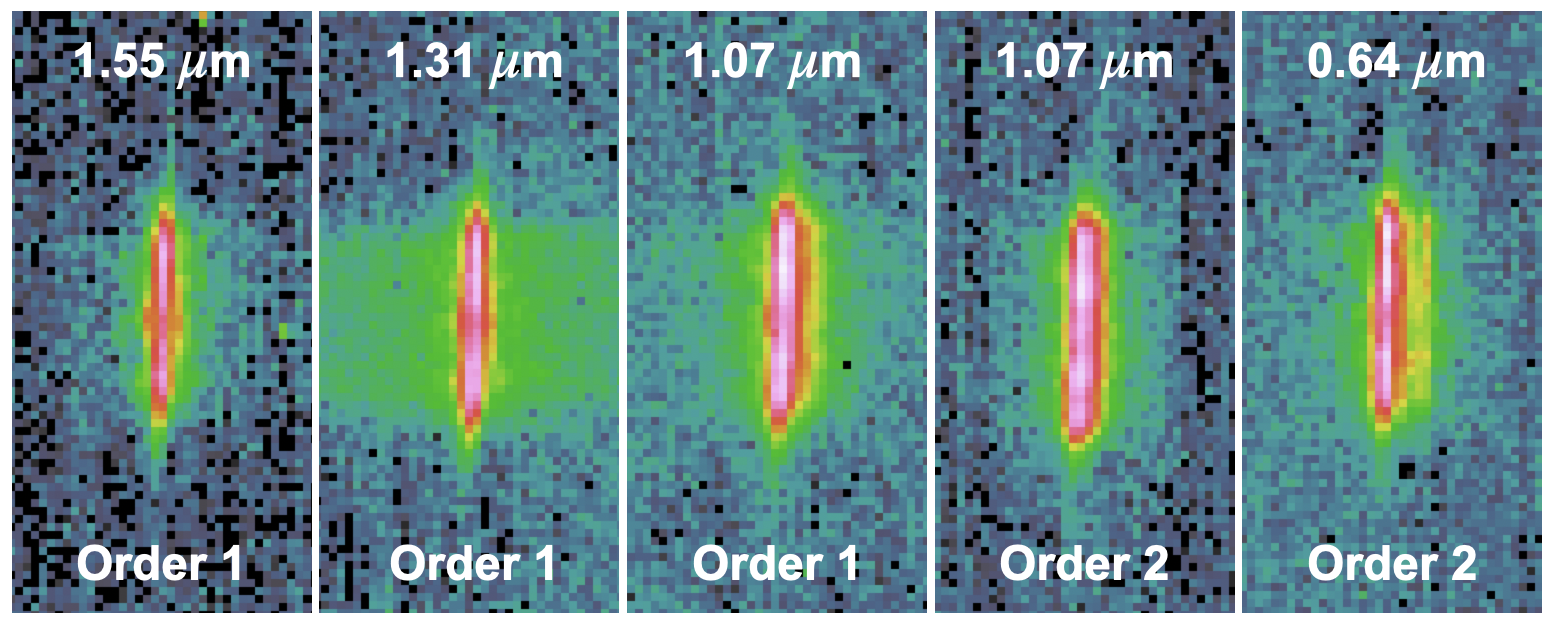}
    \caption{Point spread functions obtained at CV3 from monochromatic LED sources at 1.55\,$\mu$m, 1.31\,$\mu$m, 1.06\,$\mu$m and 0.64\,$\mu$m in both orders 1 and 2. The defocused profile along the cross-dispersion axis ($y$-axis) is roughly 23 pixels while the line resolution along the wavelength dispersion axis ($x$-axis) is (from left to right): 1.8, 2.0, 2.3, 1.8 and 2.6 pixels (FWHM). The monochromatic tilt is about 1.9\degree.}
    \label{fig:monopsf}
\end{figure}

Since NIRISS has no internal lamp for wavelength calibration, we observed four monochromatic sources (LEDs) at CV3 to check for consistency of the wavelength solution with our Code V optics model. We also used the filter cutoff near 2.4\,$\mu$m of a GR700XD + F277W (a wide-blocking filter centered at 2.77~$\mu$m) observation as a fifth wavelength anchor. 
The model and measured spectral dispersions agreed to within 1\% (see Tab.~\ref{tab:tracebounds}) at $0.96\pm0.02$\,nm/pixel in the first order and $0.467\pm2\%$\,nm/pixel in the second order. Figure~\ref{fig:monopsf} shows the PSF at four monochromatic wavelengths obtained at CV3 with SOSS. The height of each PSF is consistent with 23 pixels while their widths vary between 1.8 and 2.5 pixels (FWHM). Hence the optics is close to Nyquist sample-limited.

\begin{table*}[]
    \centering
    \begin{tabular}{cccc}
    Spectral & \multicolumn{2}{c}{Wavelength Coverage} & Dispersion ($d\lambda/dx$) \\
    Order & Minimum ($\mu$m) & Maximum ($\mu$m) & (nm/pixel)\\
    \hline
    \hline
    1 & 0.843 & 2.833 & $0.96\pm2\%$ (0.9718) \\
    2 & 0.575 & 1.423 & $0.467\pm2\%$ (0.467) \\
    3 & 0.603 & 0.956 & (0.310) \\
    \end{tabular}
    \caption{Wavelength coverage and CV3-measured dispersion of the three SOSS spectral orders with model expectations in parenthesis.}
    \label{tab:tracebounds}
\end{table*}

\subsection{Trace Overlap Contamination}
The trace overlap contamination problem can be seen in Fig.~\ref{fig:firstlight} where the traces of the first and second diffraction orders overlap toward their long wavelength end (on the left side of the subarray). The overlap covers about half the trace width. \cite{darveau-bernier_atoca} quantified the error that results from transit/eclipse observations if one performs a box aperture extraction neglecting the contamination. The error depends on the spectral type (early type stars have stronger fluxes in order 2), on the transit depth, and on the planetary spectrum. For flat planet spectra, the trace overlap produces zero net error on the extracted spectrum because both orders vary identically in relative fluxes. However, as soon as the planet spectrum differs between the first and second orders, an error arises when performing box extraction. Fortunately, for the majority of studied planets, the spectrum error is less than 10\,ppm and can therefore be neglected for most applications.

Notwithstanding, \cite{darveau-bernier_atoca} implemented {\tt ATOCA} as the SOSS spectrum extraction default for the Data Management System (DMS) \texttt{extract1d} step. It performs a trace overlap decontamination by modeling and subtracting the trace for both orders. This is useful in the context of science programs other than time-series observations which require good \emph{absolute} calibration of their spectra, contrary to the planet community which generally requires accurate \emph{relative} calibration. In that case, the absolute flux error due to the contamination depends on the relative intensities of the first and second orders which can be as high as a few percent, depending on the spectral type.

\subsection{Resolving Power}
\begin{figure}
    \centering
    \includegraphics[width=\linewidth]{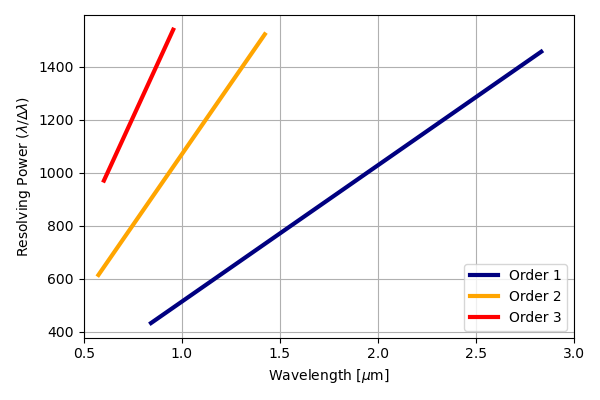}
    \caption{Optics model resolving power of the three spectral orders of SOSS, assuming two-pixel Nyquist sampling resolution elements. }
    \label{fig:resolvingpower}
\end{figure}

The requirement for the GR700XD ZnSe grism resolving power, $R = \frac{\lambda}{\Delta\lambda}$, was R=700$\pm50$ at $\lambda=1.25\,\mu$m in the first order, where $R$ is determined by equation \cite[eq. 16]{allington-smith.2005}:
\begin{equation}
    R = \frac{m \rho \lambda\, D_{pup}\, f_{col}\, cos(\beta)}{s}.
\end{equation}
Here, $m$ is the spectral order, $\rho$ is the ruling density (54.3 mm$^{-1}$), $\lambda$ is the wavelength ($1.3~\mu$m), $D_{pup}$ is the pupil diameter (39~mm), $f_{col}$ is the collimator focal ratio (f/8.8), $\beta$ is the apex angle (1.9$^\circ$) and $s$ is the slit width (2 pixels or 36\,$\mu$m) --- values in parentheses are from CV3 measurements. Using these CV3 values yields the expected resolving power, $R=646$. That expected $R$ value was confirmed by measuring the spectral trace dispersion, i.e., the derivative of the wavelength calibration (see section~\ref{sec:wavecal}) and assuming a Nyquist-limited resolution of two pixels. Using the wavelength dispersions of Tab.~\ref{tab:tracebounds},
the measured resolving power is $R=650\pm10$ at 1.25~$\mu$m in first order. 
Figure~\ref{fig:resolvingpower} also presents the resolving power for the second and third orders. 

    \subsection{Spectral Trace Profile}
In order to facilitate observations of bright targets, a weak cylindrical lens was installed just prior to the ZnSe grism along the optical path (see e.g., Fig.~\ref{fig:GR700XD}). The effect of the cylindrical lens is to defocus the 
PSF along the cross-dispersion, or spatial, direction, reducing the intensity in any one pixel, and thereby allowing for brighter stars to be observed with a given count rate per pixel.  

\begin{figure}
    \centering
    \includegraphics[width=\linewidth]{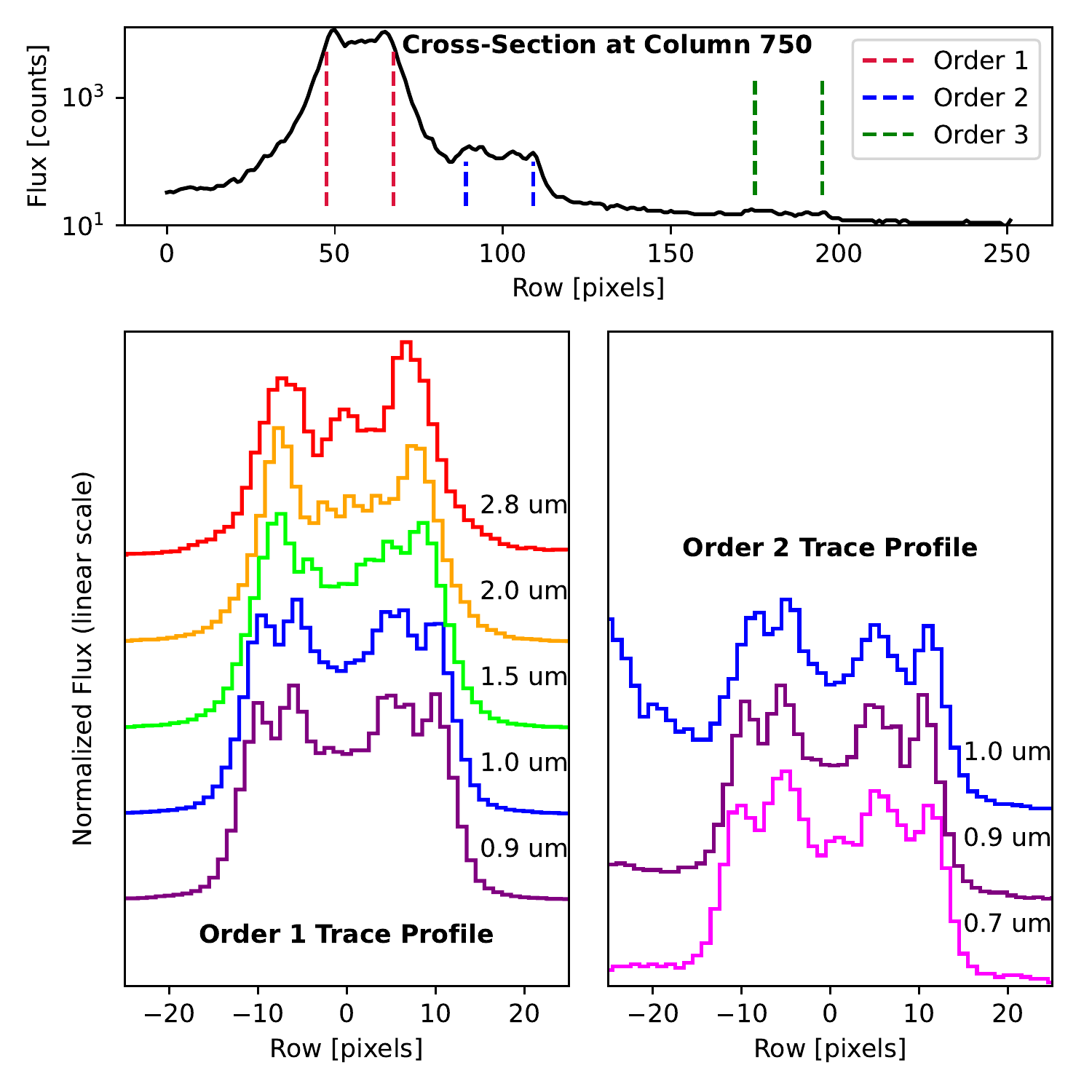}
    \caption{\emph{Top:} Full spatial profile of the SOSS spectral trace at detector column 750 ($\sim$2.1\,µm, $\sim$1.07\,µm, and $\sim$0.72\,µm for orders 1, 2, and 3; chosen so all three orders are visible) from data taken at CV3 testing. Orders 2 and 3, which are known to be particularly faint in the CV3 data due to the nature of the tungsten lamp used, 
    are nearly overwhelmed by the extended wings of the first order.
    \emph{Bottom:} Zoom in on the fine structure visible in the cores of spectral orders 1 and 2. Profiles are shown for 3 or 5 representative wavelengths to demonstrate the wavelength evolution of the trace. A general `two-horned' structure is common throughout, but significant deviations are seen with wavelength for a given order. Profiles between orders 1 and 2 at 0.9\,µm and 1.0\,µm are broadly similar although small scale differences exist probably due to different pixel sampling between the two traces.}
    \label{fig:profile}
\end{figure}

The top panel of Fig.~\ref{fig:profile} shows a representative slice along the spatial direction of the 2D SOSS spectral trace, featuring all three spectral orders. The tungsten lamp used in these CV3 tests was quite faint in the blue, and thus the second, and third spectral orders show very low intensities and are nearly overwhelmed by the extended decaying wings of the first order.  

Due to the defocusing, the Y-axis width of the PSF is not simply described by the $\lambda/D$ diffraction limit formula, although it does scale with wavelength. The trace width \textit{increases} monotonically, and near-linearly with wavelength; from 20\,pixels at $\sim$0.9\,µm to 25\,pixels at $\sim$2.8\,µm for the first order.

Additionally, the defocusing also exacerbates the high-spatial-frequency substructure, visible in the core of the trace for each spectral order (e.g., Fig.~\ref{fig:profile} bottom panel). This highly-visible substructure is due to distortions in the wavefront caused by imperfections in the optical system (e.g., the JWST optics, NIRISS and GR700XD optical systems, etc.). 
The bottom panel of Fig.~\ref{fig:profile} shows the cores of the spatial profiles for orders 1, 2, and 3 for three (or two in the case of order 3) representative wavelengths in order to demonstrate similarities and differences in the high-frequency substructure between orders, as well as the spatial evolution. All profiles display a general `two-horned' shape, roughly similar to a Gaussian with two `horns', on either side of the peak. This approximation is most valid at longer wavelengths, especially for the first order, where the substructure in the center of the profile tends to be more muted. However, there is a slow, but significant evolution with wavelength for the first order, with up to four `horns' emerging at the bluest wavelengths. Wavelength evolution is much less significant for the second and third orders, due to their limited wavelength coverages. For example, from $0.7 - 1.1$\,µm, the spatial profile of the second order remains qualitatively similar to the 1.0\,µm profile in the first order (compare the bottom right panel of Fig.~\ref{fig:profile} to the blue curve in the left-hand panel). The spatial profiles are therefore nearly identical at a given wavelength between spectral orders, as expected.

For modeling purposes, it was found that the profile of the wings for all orders and wavelengths can be described as:
\begin{equation}
    {\rm Flux} \propto 10^{-0.005 \Delta y}
\end{equation}
where Flux is in ADU/s and $\Delta y$ is in pixels.

    \subsection{Monochromatic Tilt}
\begin{figure}
    \centering
    \includegraphics[width=\linewidth]{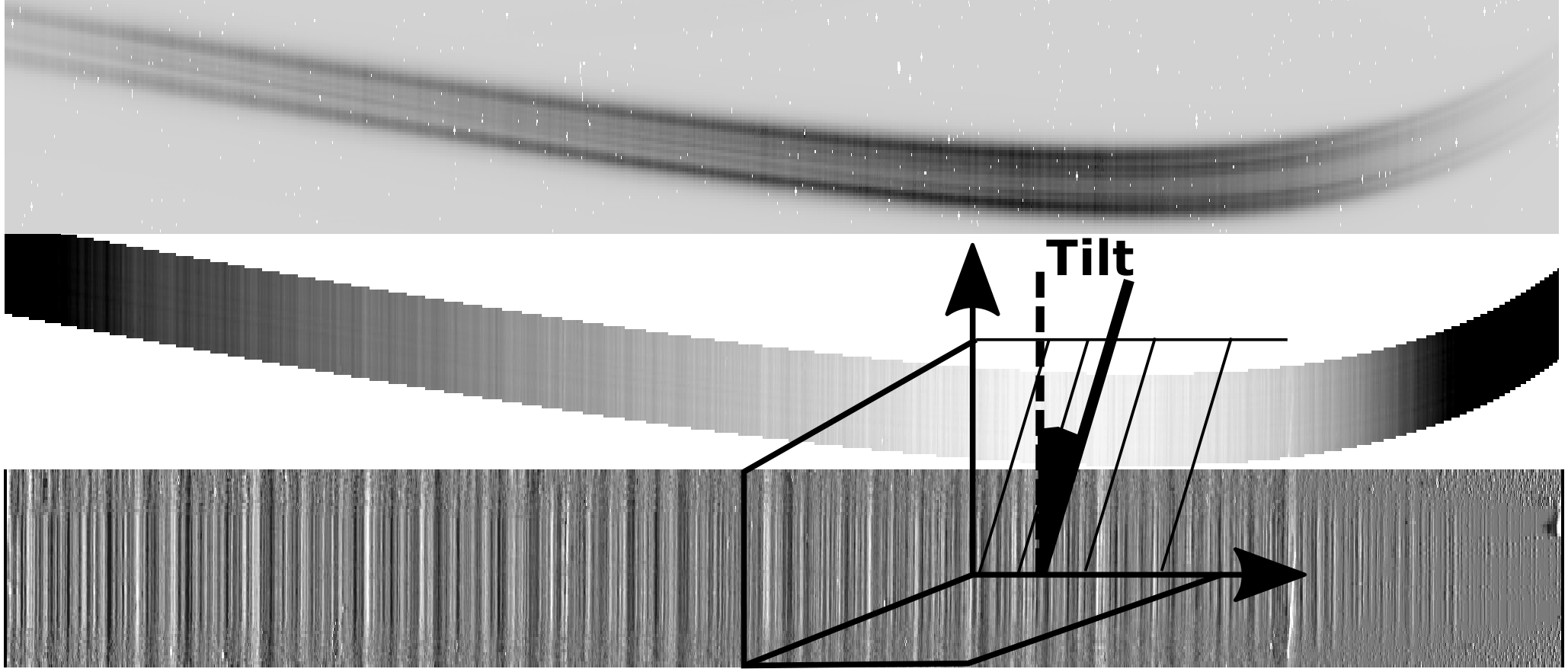}
    \caption{Monochromatic wavelengths have a small tilt relative to the detector columns. \emph{Top}: The order 1 trace of a tungsten lamp observed at CV3. 
    \emph{Middle}: The CV3 trace normalized by using a model of the trace profile. \emph{Bottom}: The CV3 trace rectified and normalized by dividing by the median spectrum. All images are stretched along the $y$-axis for clarity. The measured monochromatic tilt across order 1 varies between 0.4 and 1.2 degrees.}
    \label{fig:tilt}
\end{figure}

A small rotational offset between the cylindrical lens and the grism was purposely dialed into the GR700XD element. The intent was to spread an emission line (or any spectral feature) at a slight angle with respect to the detector columns. Since the spread is over $\sim$20 pixels, this allows sub-pixel sampling of a line in order to recover better spectral resolution. This so-called monochromatic tilt is not constant because it is also influenced by field distortion. It ranges between 0.4 and 1.2 degrees for order 1. Note that the general curvature of the cross-dispersed traces (determined by the glass refractive index dispersion) has no bearing on the monochromatic tilt; they are not connected, as is shown in Fig.~\ref{fig:tilt} where tilt remains after the trace was rectified.

    \subsection{Optical Ghosts \label{sec:ghosts}}
\begin{figure}
    \centering
    \includegraphics[width=\linewidth]{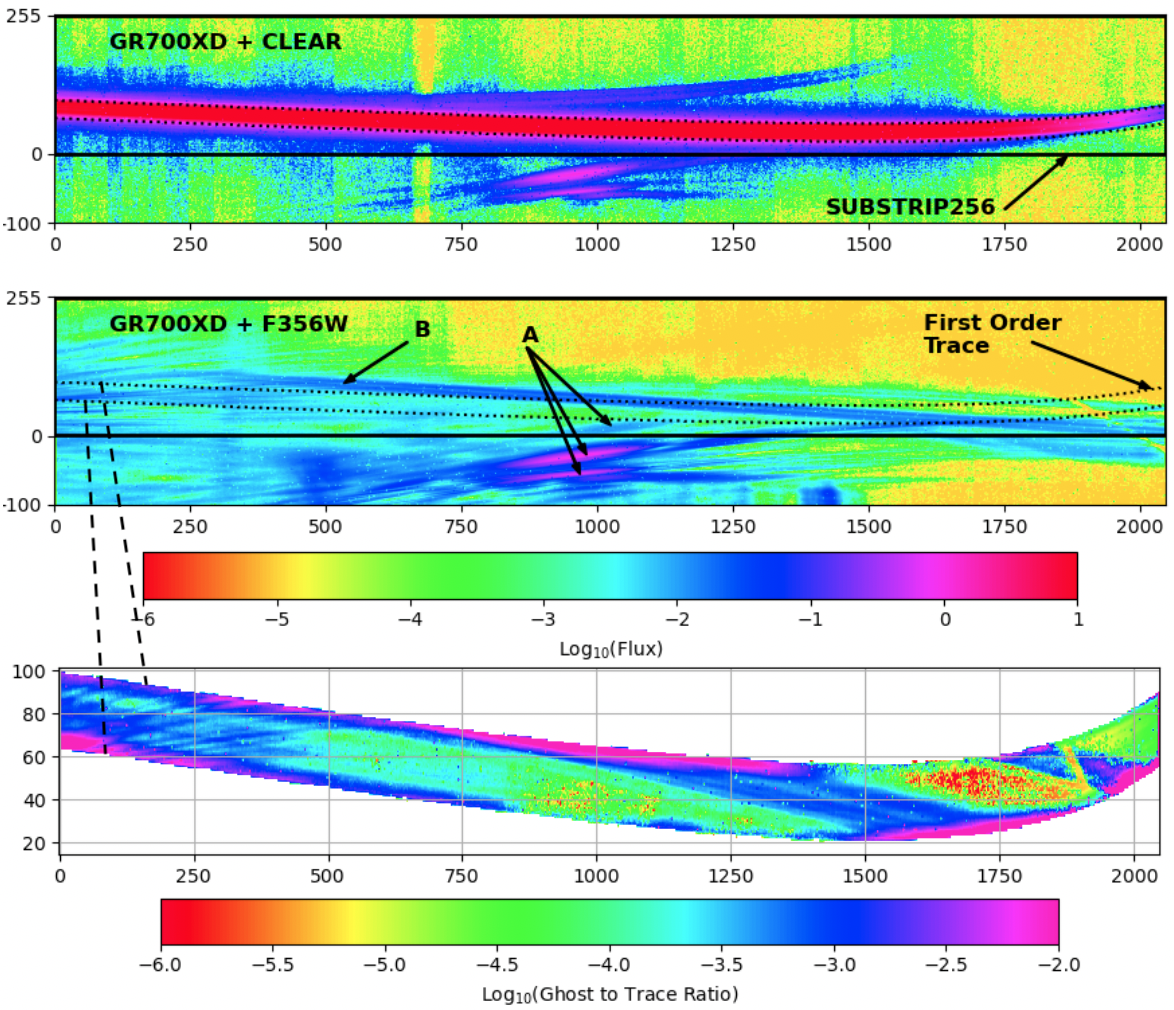}
    \caption{Ghosts seen at CV3 in the GR700XD$+$CLEAR science configuration (top) and in the GR700XD$+$F356W engineering test (middle). The ratio of the two images quantifies the relative ghost-to-trace intensity (bottom). It is calculated within the 35-pixel aperture of order 1. Two ghost sources are identified: (A) an order 0 reflection in the camera Three-Mirror Assembly is found close to the first order trace apex; (B) the diffraction pattern caused by the GR700XD square aperture mask overlaps with part of the first order trace at a level of $\sim 10^{-3}$.}
    \label{fig:ghosts}
\end{figure}

High-intensity tungsten lamp integrations were obtained to highlight ghosts, all of which can be traced back to specific optical paths based on our as-built optics and mechanical models. Figure~\ref{fig:ghosts} shows the various ghosts in two exposures, the GR700XD$+$CLEAR configuration (top) and the GR700XD$+$F356W (a wide-band filter centered at 3.56~$\mu$m) configuration (middle). Exposures with the GR700XD coupled with the F277W or F444W (centered at 4.44~$\mu$m) showed similar features. The most concerning ghost is a long and straight structure, about 25 pixels wide, whose double-horn internal structure resembles that of the science traces (ghost B in Figure~\ref{fig:ghosts}). This ghost overlaps with the SOSS order 1 trace ($1.2\leq\lambda\leq1.8$\,$\mu$m) at a level of $\sim10^{-3}$ with respect to the trace. It is caused by diffraction through the square aperture mask of the GR700XD. Another concerning ghost occurs from the zeroth order being diffused off the housing wall of the camera Three-Mirror Assembly, between mirrors 2 and 3, reflecting back to the detector (ghost A in Figure~\ref{fig:ghosts}). It lands in about three different elongated spots within 100 pixels of the SOSS order 1 trace (corresponding to $1.7\leq\lambda\leq2.1$\,$\mu$m). Fortunately, most of these ghosts fall off the science trace but they could bias slightly any background measurement using pixels in the wing of the order 1 trace, below its apex. There is potential to use the GR700XD $+$ F277W calibration observation to subtract these ghosts (see Section~\ref{sec:opsconcept}). Inspection of GR700XD $+$ F277W Commissioning images can not detect ghosts at a level of $\geq5\times10^{-3}$.

\section{Operation Concepts} \label{sec:opsconcept}

SOSS science observations will typically last three to six hours for transit and secondary eclipse spectroscopy and tens of hours for phase curves. They are performed in staring mode with the target at a fixed detector position without dithering. This is to minimize differential detector systematic errors introduced, for example, by the flat field, non-linearity, bad pixels, or intra-pixel response. Each observation consists in a single exposure containing a large number of up-the-ramp integrations, each sampled by a number of reads or groups. Typical target magnitudes should be within the range $6\leq J \leq 15$. Target acquisition through an imaging filter is first performed to position the target at the acquisition spot, then the wheels are rotated to insert the GR700XD and CLEAR optics, which effectively projects the SOSS traces near the top edge of the detector where one of the three possible subarrays reads out the detector. At the end of the science exposure, an optional short calibration sequence using the F277W filter in combination with the GR700XD is possible to calibrate the trace profile of the first spectral order.

    \subsection{Science Subarrays}
SOSS offers a choice of three readout modes (refer to Fig.~\ref{fig:layout} --- subarrays are represented as different 
rectangles). The nominal SUBSTRIP256, a 256$\times$2048 subarray, is designed to capture the core and a good part of the pixels in the wings of the diffraction orders 1, 2, and 3. It is positioned along the slow-readout direction of the detector, up to its edges on the amplifier 1 side. The position of this subarray was chosen so as to include the four-pixel wide stripe of reference pixels. SUBSTRIP96, a 96$\times$2048 subarray, is designed to observe brighter targets without saturating. It includes only the first diffraction order trace and has no access to the top rows of reference pixels. The bottom row of SUBSTRIP96 is offset by 10 pixels on the detector in comparison with SUBSTRIP256. Both subarrays are read out using amplifier 1. The third readout mode, FULL, uses the 4-amplifier full-frame readout mode. Its larger frame time makes it less observing-time efficient, but for faint targets, it can potentially help with 1/$f$ noise subtraction and offers the opportunity to observe a field target simultaneously.

    \subsection{Target Acquisition}
Guiding with the Fine Guidance Sensor (FGS) and target acquisition are mandatory for all SOSS observations. Imaging and centering the target are necessary to ensure that the spectral traces are projected on the detector at a repeatable position, primarily to facilitate wavelength calibration. Also, having an archive of SOSS sequences at the same position should allow for the construction of a better trace profile and ultimately calibrate out the inter-order contamination of SOSS.

The Acquisition Spot is the detector position  where the target acquisition algorithm attempts to center the science target: $x=1955$, $y=1199$ (DMS coordinates). To achieve the 1/15 of a pixel acquisition accuracy as well as reject cosmic ray events, exposures at three dither positions are obtained with the SUBTASOSS subarray, a $64\times64$ pixel subarray located at position $1202 \leq x \leq 1265$ and $392 \leq y \leq 455$ with a readout frame time of 50.16\,msec. The acquisition is obtained through the F480M + CLEARP\footnote{The 4.8~$\mu$ medium (5\%) blocking filter (F480M) is in the filter wheel and the CLEARP element is an open hole in the pupil wheel.} filters (faint targets) or through the F480M + NRMMASK\footnote{NRMMASK is a pupil mask with 7 hexagonal openings matching 7 JWST mirror segments normally used to perform non-redundant masking interferometry, but, in the context of target acquisition, used to attenuate the light by a factor of 7.} (bright targets). The faint target choice can yield a peak SNR $\geq 30$ without saturating for stars with $6.5\leq$ F480M $\leq 12.5$ while the bright target choice achieves the same on stars with $3.0 \leq$ F480M $\leq 8.5$, depending on the exact choice of readouts selected (3 to 19 in steps of 2).

    \subsection{Detector Readout}
The integration time when performing Correlated Double Sampling (CDS) is set by the frame time as well as the number of detector readouts (($N_{\rm group}-1) \times t_{\rm frame}$). For NIRISS, two readout schemes are allowed: 1) {\tt NIS} --- a {\em group} consists of the average of four detector readouts to save storage space onboard; 2) {\tt NISRAPID} --- a {\em group} consists of a single readout, i.e., all readouts are saved. We recommend the use of {\tt NISRAPID} for all SOSS exposures.

The readout frame time is governed by the equation:
\begin{equation}
    t_{\rm frame} = t_{\rm pixel} \times (({\rm cols} / {\rm amps} + 12) \times ({\rm rows} + 2) + {\rm pixpad})  
\end{equation}
where $t_{\rm pixel} = 10\,\mu$s and rows\,$= 2048$. For SUBSTRIP96/SUBSTRIP256, cols\,$= 96/256$, amps\,$=1$ and pixpad\,$ = 0$ while for FULL, cols\,$ = 2048$, amps\,$=4$ and pixpad\,$ = 1$.

Table~\ref{tab:saturation} lists the readout times associated with the SUBSTRIP96, SUBSTRIP256, and FULL subarrays.

    \subsection{The GR700XD + F277W Trace Profile Calibration}
The Astronomer Proposal Tool (APT) allows for an optional calibration in the SOSS mode, i.e., an observation using the F277W filter and the GR700XD element. This is executed immediately after the normal, GR700XD + CLEAR, science time-series to ensure a calibration at the specific Pupil Wheel motor step position. This GR700XD+F277W calibration entirely suppresses the second and third-order spectral traces and cuts the first order below $\sim2.4~\mu$m.

There are two reasons for this calibration. First, it helps mitigate the cross-order contamination problem by providing a measurement of the first-order trace in isolation to allow the construction of a trace profile function. In turn, that profile can be scaled and subtracted from the science observation to retrieve the second-order trace profile function. Second, it potentially helps characterize optical ghosts appearing underneath the science traces (c.f.\ Sect.~\ref{sec:ghosts}) and helps identify field contaminants through their order 0 signature. We caution that the GR700XD+F277W observation, however, does not help mitigating the contamination by field stars.

\begin{figure}
    \centering
    \includegraphics[width=\linewidth]{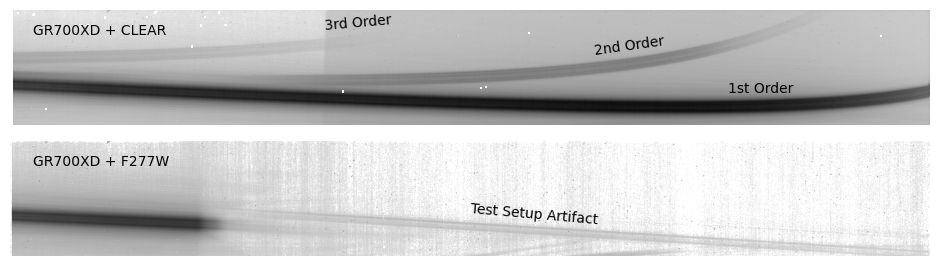}
    \caption{Side by side CV3 images of the Science configuration GR700XD~$+$~CLEAR (top) compared with Calibration configuration GR700XD~$+$~F277W (bottom). The F277W calibration is performed on the science target after the science exposures to obtain a measurement of the first-order trace profile, free of second-order contamination.}
    \label{fig:f277}
\end{figure}

    \subsection{Saturation Limiting Magnitudes}
Given the digital saturation limit of 65536\,ADU for our 16-bit analog-to-digital converters and the image bias of roughly 12\,000 -- 15\,000\,adu, the effective well depth of the NIRISS detector is $\sim$50\,000\,ADU, or $\sim$80\,000\,e$^-$ for a gain of $\sim1.6$\,e$^-$\,ADU. Roughly half of the pixels reach digital saturation before large $\geq$10\% non-linearity occurs. For computing saturation limiting magnitudes, we adopted a level of 90\% of the digital saturation, or 45\,000\,ADU (72\,000\,e$^-$). Note that the Brighter-Fatter Effect could dictate a more conservative 30\,000\,ADU limit for these calculations (See Sec.~\ref{sec:bfe}). We used our SOSS simulator (See Sec.~\ref{sec:idtsoss}) to generate images of targets with known magnitudes to measure the effective count rates in the brightest pixels of the trace and scaled the measurements to calculate the Vega limiting magnitudes given in Table~\ref{tab:saturation}. These figures are dependent on the exact trace profile shape, which in turn depends on the telescope optics wavefront, so our estimates are uncertain to about $\sim$0.1--0.2\,mag. We repeated these measurements on stars of two spectral types (M3V and F2V) and obtained the same results to $\leq$0.05\,mag. Finally, we added 0.15\,mag to account for the 10--20\% better throughput measured during Commissioning.

\begin{figure}
    \centering
    \includegraphics[width=\linewidth]{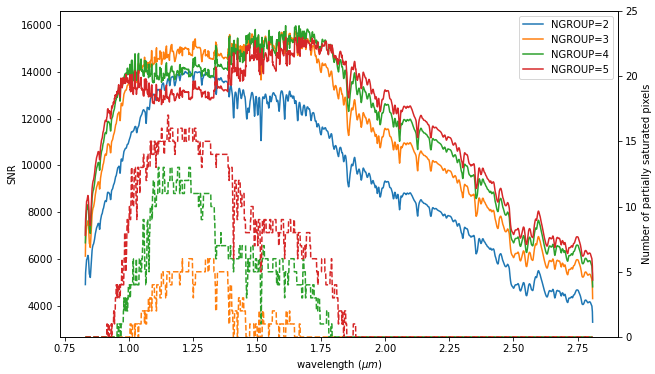}
    \includegraphics[width=\linewidth]{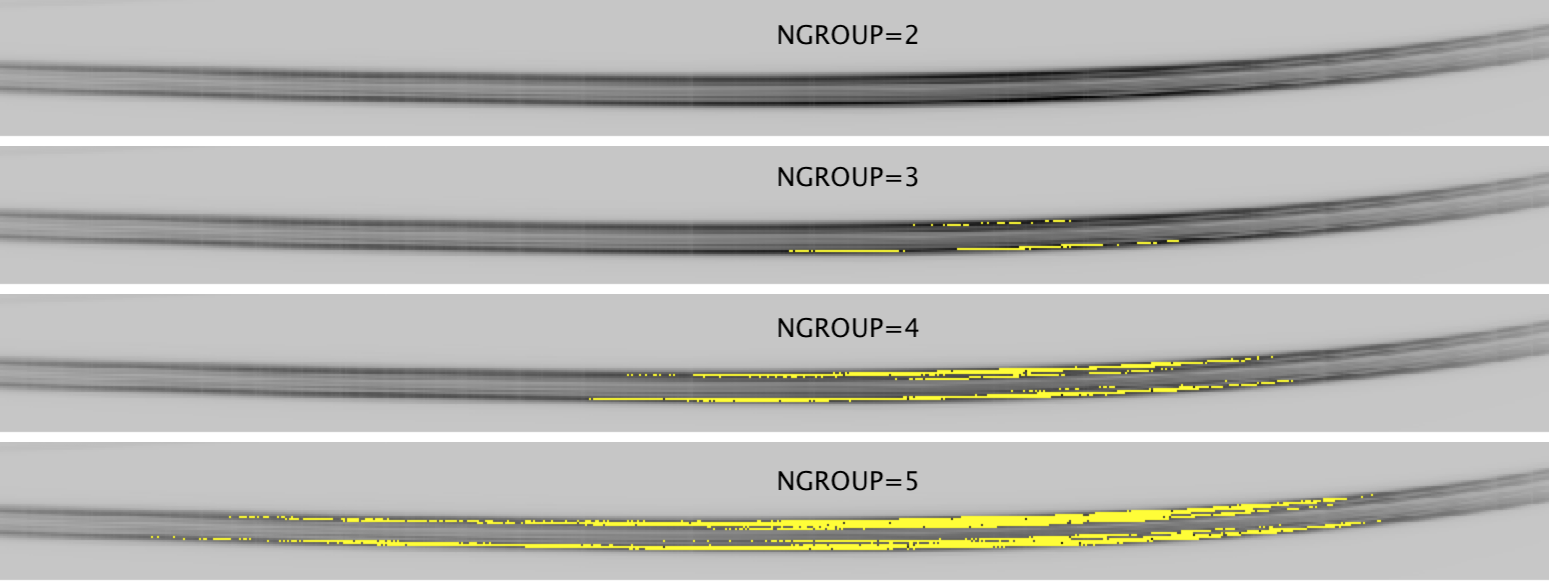}
    \caption{Extracted spectra and signal-to-noise ratios are displayed in the top panel of the simulation of GJ\,357 using the {\tt IDT-SOSS} simulator (See Sec.~\ref{sec:idtsoss}) whose detector images are shown at the bottom. Each curve or image is for integrations with different numbers of group readouts between NGROUP=2 and 5. The onset of saturation occurs at NGROUP=3 in the horns of the trace and spreads to a wide wavelength range by NGROUP=5, while remaining limited to the horns. Unsaturated pixels in the trace core still convey useful information. These plots show that saturation limiting magnitudes should be regarded as an indicator rather than a strict limit.}
    \label{fig:partialsat}
\end{figure}

SOSS saturation first occurs near the peak of the grism blaze function, at 1.2\,$\mu$m, in the two peaks of the trace profile, then at the center of the trace profile (about half the peak) --- see Fig.~\ref{fig:partialsat}. This means that roughly half of the pixels at any given wavelength can be exposed for twice longer before the full saturation of all pixels at that wavelength.

\begin{table}[]
    \centering
    \begin{tabular}{cccc}
    Subarray & $t_{frame}$ & \multicolumn{2}{c}{$J$-band saturation limit} \\
             & (sec)       & $N_{group}=1$ & $N_{group}=2$ \\
    \hline \hline
    SUBSTRIP256 & 5.494 & 7.69 & 8.44 \\
    SUBSTRIP96  & 2.214 & 6.70 & 7.45 \\
    FULL     & 10.74201 & 8.41 & 9.16 \\
    \end{tabular}
    \caption{Subarray frame time and $J$-band saturation limits from ground-based characterization and simulations, which have a 5\% spread depending on the spectral type. Flight performance indicated a 10--20\% better throughput compared to predictions, we have thus added 0.15\,mag to the limiting magnitudes in this table. Commissioning on HAT-P-14b showed that observing with mild saturation is not detrimental.}
    \label{tab:saturation}
\end{table}

Moreover, if care is taken to use partial ramps to reconstruct the saturated wavelengths, a strategy that saturates part of the pixels or spectral range can potentially produce $\sim$30--40\% better SNR overall. This particular strategy was tested at Commissioning on HAT-P-14b and results are presented in Sec.~\ref{sec:noiseperformance}.

\subsection{Ngroup = 1 Observations and kTC Noise}
Raw near-infrared detector readouts harbor a large picture frame pattern across the array. Usually, this pattern cancels out when subtracting two consecutive reads in an integration. However, to push observations to bright targets, the APT allows the use of NGROUP=1, i.e., the detector is reset and pixels are read out exactly once.  This is used when saturation would already occur at the second read, preventing the use of CDS. Without two consecutive readouts, the detector picture frame pattern can be removed only by using a so-called super bias built from a large number of dark exposures. However, an NGROUP=1 image also suffers from the kTC noise, a white noise inherent to resetting the detector which affects every pixel. At a level of $\sim$100\,e-, the kTC noise is several times larger than the regular readout noise. But the kTC noise can be tolerated if NGROUP=1 enables observations of a star that would otherwise saturate and whose error budget is, in any event, photon-noise dominated.

\section{Detector Characterization} \label{sec:detector_characterization}

    \subsection{Detector Temperature and Signal Stability}
\begin{figure}
    \centering
    \includegraphics[width=\linewidth]{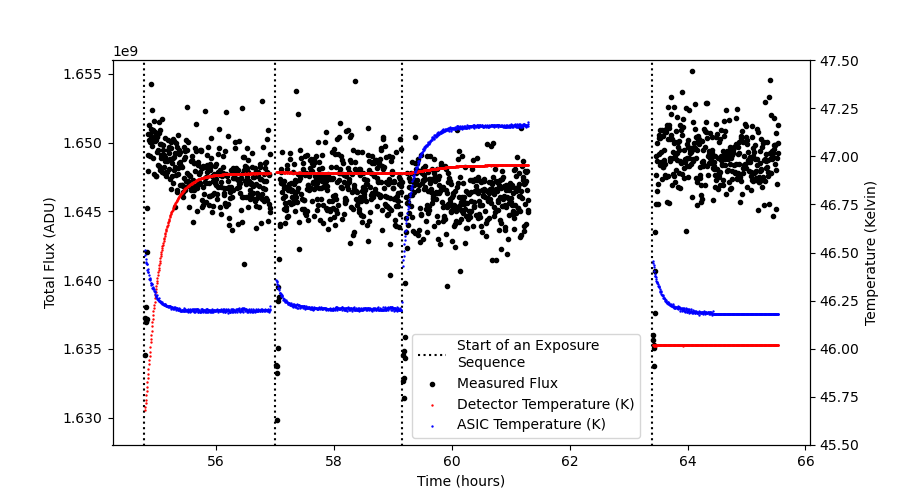}
    \caption{Detector stability through four sequences of FULL frame exposures when uniformly illuminated. The black points are measurements of the median flux on the array while the red and blue curves are, respectively, the detector and ASIC temperatures. All four sequences display a sharp increasing flux trend with time ($\sim$1\%) lasting roughly 10 integrations, after which the flux plateaus. The rightmost sequence starting at $\sim$63.5 hours is the only sequence where the detector temperature control loop was closed and coincides with a stable flux level. The first sequence on the left has the strongest flux variation and it anti-correlates with the detector temperature. Sequences 2 and 3 display relatively flat flux levels and stable detector temperatures while the ASIC temperature varies, suggesting that flux measurements are robust against ASIC temperature variations.}
    \label{fig:stability}
\end{figure}

During detector characterization, four 1-hour-long sequences of uniformly illuminated flat fields were gathered to characterize the detector response stability and look for any ramp effect at the start of an exposure. This test was conducted by the NIRISS contractor, Honeywell, on the actual NIRISS detector and Spare Flight system (Fig.~\ref{fig:stability}). The NIRISS detector is clocked by the ASIC controller and both have temperature sensors and separate temperature control loops. A short duration ($\sim 10$ integrations) rising trend of about 1\% occurs at the start of every sequence for less than 10 integrations but the flux then stabilizes at a constant level with deviations at the $\sim 10^{-3}$ level. The four sequences differ in how the ASIC and detector temperatures were controlled. The first sequence left both ASIC and detector temperatures uncontrolled which resulted in a downward trend for the measured flux. The flux in the second sequence was mostly stable while both the detector and ASIC temperatures were in an open loop but mostly stable. The third sequence shows that an upward-trending ASIC temperature did not affect the measured flux. The fourth sequence was obtained with the detector temperature stabilized with the control loop closed while the ASIC temperature was left varying, and again, the measured flux was stable. These sequences suggest that the detector temperature is the sole parameter correlating with measured flux ($-1.4\times10^{-3}$/K). The ASIC temperature does not affect our measured flux. In flight, the NIRISS detector is controlled at a stable temperature with milli-Kelvin accuracy to prevent temperature-related flux drifts.

    \subsection{1/f Noise}
The 1/$f$ noise is a time-correlated noise characterized by vertical stripes in SOSS images at a level of a few, to a few tens of electrons. This arises because pixels are read sequentially at a fixed $10^5$\,Hz rate while the reference voltage against which each pixel's voltage is compared fluctuates with a 1/$f$ power spectrum \citep[for more details, see][]{rauscher.2014}. The 1/$f$ noise is neither fixed in time nor in detector position, so it cannot be calibrated out using, for instance, reference darks. This is why a four-pixel wide row of light-insensitive pixels (i.e., the reference pixels) are present at the edges of the H2RG detectors. Unfortunately, they do not sample the 1/$f$ noise at a high enough cadence to allow one to fully eliminate the effect. For SOSS, we encourage the use of light-sensitive pixels, as long as pixels receiving astrophysical signals are properly masked, to model the 1/$f$ noise and subtract it on a per-column basis. 

    \subsubsection{Correcting 1/f at the Group Level \label{sec:1overfcorrection}}

We developed a method to correct for 1/f noise at the group level (on each raw image) rather than after having fit the slope (at the integration level). This is because, conceptually, since the 1/$f$ noise is the last source of detector noise imprinted on the image, it should be the first removed. We first build a high SNR stack of the raw groups from all integrations available in a TSO, then subtract it from each group in each integration to effectively suppress most of the constant astrophysical signal (spectral traces, zodiacal background) as well as the picture-frame (superbias) signal. This leaves photon noise, readout noise and 1/f structures in the residuals map. We further mask out the pixels with large photon noise where the spectral traces land using a 30-pixel wide mask. This effectively removes any time-varying small residual signal not captured when subtracting the deep stack, such as transit events. Finally, assuming that the 1/f noise is constant across a column, we compute the average (noise-weighted) of each column of each group and subtract it from the raw images.

\begin{figure*}
    \centering
    \includegraphics[width=\linewidth]{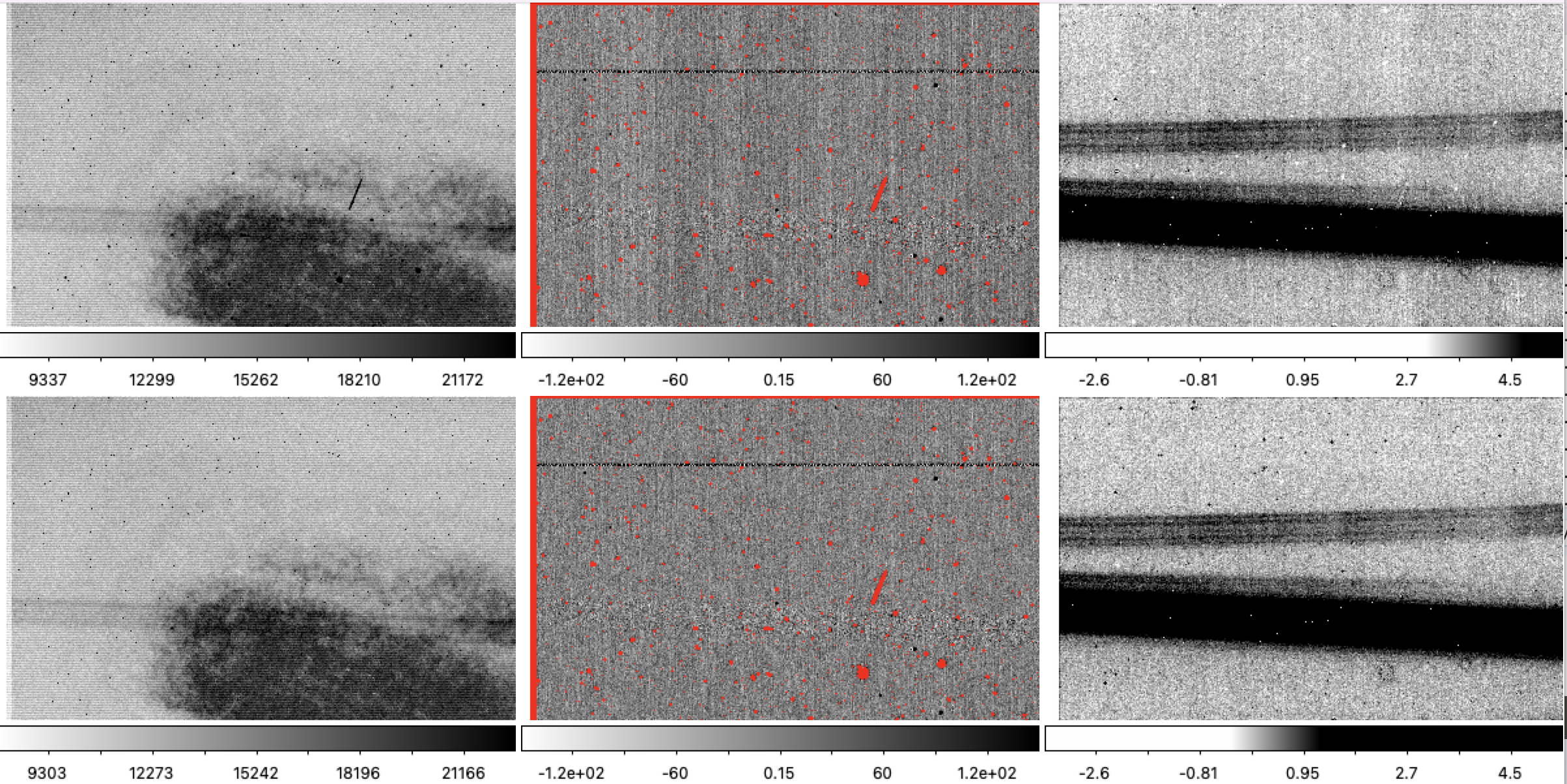}
    \caption{Close up of the 390$\times$256 left-most pixels of a SOSS SUBSTRIP256 Trappist-1 TSO. On the top-left is a single read (i.e. group 18 of the 56th integration) and below (bottom-left) is the median group 18 for all integrations available in the TSO. In the central panels, the median group 18 stack is subtracted off integration 56's group 18 (central-top), the 1/f noise fitted as a constant in each column the subtraction is the central-lower panel. At right, we compare the fitted slopes between ({\tt rateints.fits} the default DMS output (top right) and our custom output (lower-right).
    Note the presence of pair or noisier rows in the signal-subtracted central panel. These are discussed in the text.}
    \label{fig:1overf}
\end{figure*}

Two consecutive rows of noisier pixels ($\sigma \sim 100$~adu) can be seen in the central panels of figure~\ref{fig:1overf}. Their position moves one row down with every new integration, leaving the sub-array entirely at integration 257. It was checked that these rows do not roll back after 512 or 2048 integrations in an exposure. These noisier rows appear linked to the detector read-out strategy, in particular to the full detector reset performed after every sub-array integration, before sub-array reset, to prevent charge bleeding from outside of the sub-array.

    \subsection{Pixel Response }
    \label{sec:bfe}
\begin{figure*}
    \centering
    \includegraphics[width=\linewidth]{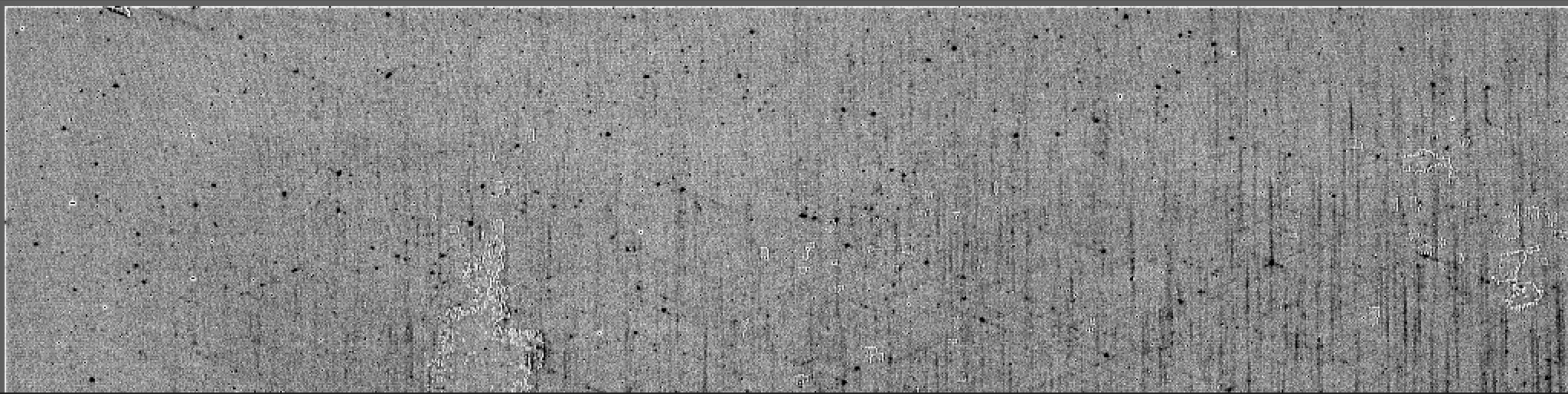}
    \caption{Post Commissioning SOSS mode flat field in the SUBSTRIP256 subarray (left 1024 columns shown). Epoxy void patches are seen as brighter pixels forming closed loops while cross-hatching are the darker features along 3 axes: vertical and both diagonals forming Xs towards the bottom. Groups of dark pixels are generally bad pixels or much less light sensitive pixels}
    \label{fig:flatfield}
\end{figure*}

The flat field released for SOSS at the end of Commissioning strongly relies on measurements obtained during ground testing at the CV3 campaign. We did not observe a strong change in the flat field properties of the imaging filters from 0.9\,$\mu$m (F090W) to 2.77\,$\mu$m (F277W) so we selected the flat at 1.15\,$\mu$m in filter F115W, being near the peak of the order 1 response, as the best approximation to the SOSS mode flat field given that we lack monochromatic flat field measurements. The flat field was obtained in Imaging mode, a mode for which the four occulting spots that were engraved into the Pick Off Mirror can be seen. They disappear in the dispersed images of SOSS. So, for the purpose of SOSS flat field construction, the pixels in these holes were replaced by those in the dispersed flats gathered through the GR150 grisms. That step was also applied to various pixels where dust on the Pick Off Mirror produced features that seem to have moved slightly after optics alignment. Onboard lamps, a legacy of the TFI, can illuminate the detector but in a far from uniform fashion ill-suited for flat calibration. A ratio of lamp flats obtained on the sky was compared to CV3 lamp flats and this ratio was applied to the F115W flats to provide an update of the pixel flats. Finally, a low-frequency flat correction was applied using the flux variations measured when dithering stars over the field of view in Imaging mode. The resulting flat field produced after Commissioning is shown in Figure~\ref{fig:flatfield}. We note that since our flat neglects any potential wavelength dependency, any residual error is fortunately corrected at a later stage: during absolute flux calibration ({\tt PhotomStep} in the DMS). We caution, though, since the spectral traces can move by several pixels between science targets, the absolute flux calibration may not entirely correct for the flat's wavelength dependency. This is less a concern for time-series observations interested in relative flux variations but should be noted for projects relying on absolute flux calibration. Ultimately, the optimal flat fielding approach is to develop a model of the spectral traces profile at each wavelength and couple this to actual observed traces to retrieve the flat field response.

Several detector artifacts are apparent in the flats, most obvious are the epoxy voids and cross-hatching. The process of filling with epoxy the space between the silicon ROIC and the HgCdTe detector to strengthen its mechanical structure can leave empty holes which we see as clumps of brighter pixels. It was found that the gain (e-/ADU) is reduced for the pixels in a void, making them appear brighter in the flats. Epoxy void pixels are not more sensitive than regular pixels. Cross-hatching is potentially more problematic for accurate photometry. The layer of HgCdTe that grows from the beam epitaxy process has a tendency to adopt a crystalline structure, forming nanometer-sized bumps that can shadow light and produce an x-shaped response to flux at the sub-pixel level  \citep[see][]{schlawin.2021}. We refer the reader to \cite{rauscher.2015} for a review of all detector artifacts based on the NIRSpec detectors which are of the same family as those in NIRISS.

\subsection{Intra Pixel Response Concerns}

For regular pixels of HxRG detectors, the sub-pixel response has been shown to be uniform when scanning their response with a small laser beam  \citep[see][]{barron.2007,hardy.2014}). However, photo-electrons can diffuse to neighboring pixels if a photon is incident in the periphery of a pixel. That said, micron-sized defects do exist and do change the response at the sub-pixel level of any affected pixels. More importantly, pixels in cross-hatched regions indeed suffer from a non-uniform sub-pixel response \citep{shapiro.2019} with residuals at the $\sim2\%$ level after flat fielding. For the SOSS mode, the contrasted structures in the spectral trace profile, if combined with tiny image motion, may affect the photometric accuracy for pixels in cross-hatched regions. The SOSS SUBSTRIP256 subarray has mild cross-hatching for columns 500 to 1400 (1.5--2.3 $\mu$m in order 1), as seen in Fig.~\ref{fig:flatfield}. Finally, it was found at Commissioning that ground and Flight Imaging flats differed by several percent in regions with severe cross-hatching, likely due to a slight converging beam incidence angle (F$/\#$) mismatch between the Ground Optics Simulator and Flight Optics.

\subsection{The Brighter-Fatter Effect and Inter-Pixel Capacitance}

The Brighter-Fatter Effect (BFE) is a form of charge migration where the repulsive Coulomb force experienced by freshly produced electrons is stronger near bright pixels, effectively leading them away to neighboring, lower flux, pixels \citep{antilogus.2014}. The BFE for the NIRISS detector kicks in at $\sim$30\,000\,ADU$^-$, roughly half the full-well depth \citep{sivaramakrishnan.2023}. Since non-linearity is characterized using uniform illumination where the BFE between adjacent pixels more or less cancels out, for actual observations in the presence of an astrophysical signal, the non-linearity correction for BFE affected pixels is underestimated for bright central pixels and overestimated for adjacent fainter pixels, thus affecting the photometry.

The Inter-Pixel Capacitance (IPC) is an electric capacitive coupling that produces a ghost signal in pixels neighboring a bright pixel \citep{moore.2004}. For the NIRISS detector, a bright pixel has 2.4\% of its flux induced in the four adjacent pixels. Technically, the IPC is a convolution of the image by a small IPC kernel. There is a DMS pipeline step to correct for this but it is not used for the SOSS TSO pipeline. Worth noting for SOSS, the detector void pixels have a different IPC than the non-void pixels.

\section{In-Flight SOSS Mode Characterization} \label{sec:flight}
    
    \subsection{Flux Calibration} \label{sec:fluxcal}
As part of the spectral reduction, the measured signal rates need to be transformed into physical units.  In the JWST pipeline, the output units selected are $F_\nu$ in MJy.  To carry out this conversion a photometric standard star, BD$+$60$^\circ$1753, was observed using the SOSS mode in the commissioning program APT\,1091.  This star was selected from the Calspec standard star list \citep{bohlin.2014}\footnote{For details see https://www.stsci.edu/hst/inst rumentation/reference-data-for-calibration-and-tools/astronomical-catalogs/calspec}, because it is of suitable brightness for SOSS observations and has good visibility over most of the year. The star is of spectral type A0mA1V and has been vetted for variability.

\begin{figure}
    \centering
    \includegraphics[width=\linewidth]{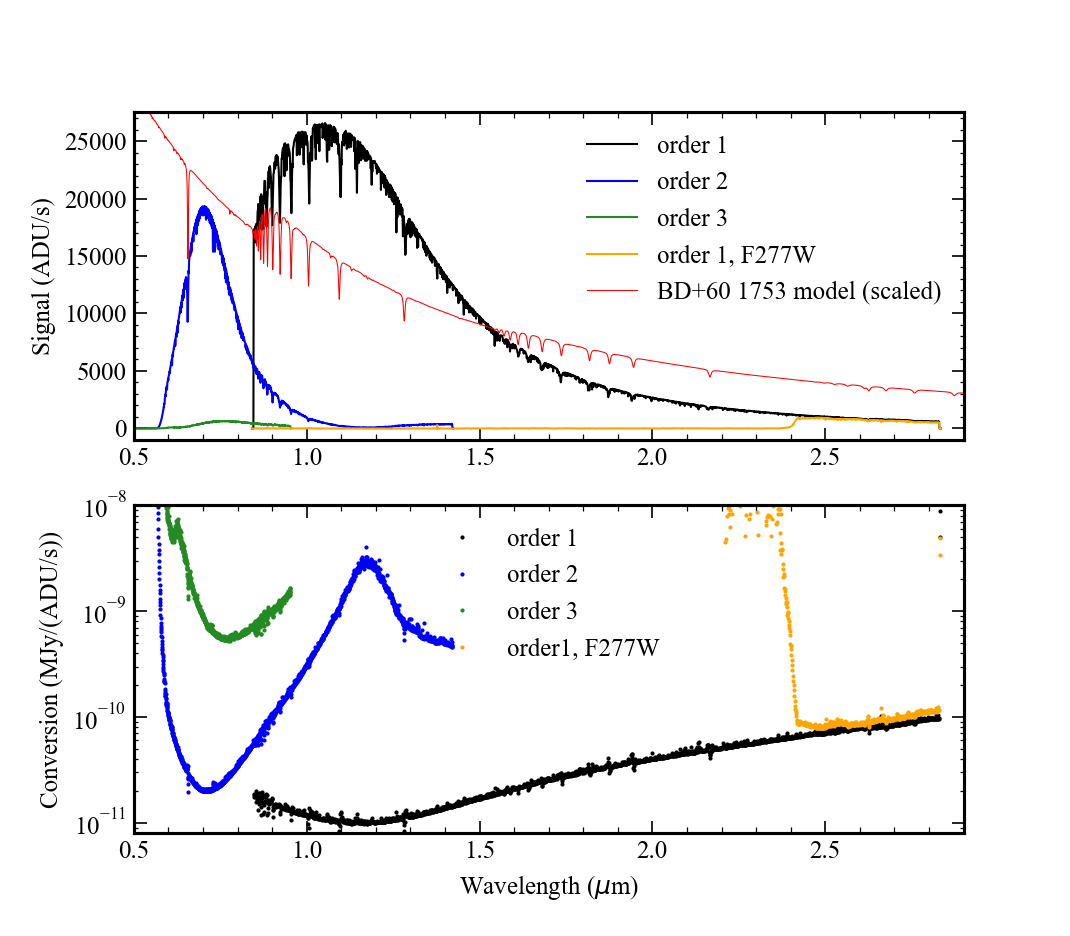}
    \caption{Illustration of the photometric calibration process.  In the upper panel is shown the extracted spectra for the target BD$+$60$^\circ$1753 in instrumental units for each of the orders.  The theoretical spectrum of the star in $F_\nu$ is also shown in the upper panel, smoothed to a uniform resolution of 1000 and scaled arbitrarily to fit within the plot.  In the lower plot the output photometric calibration functions derived from the spectra in the upper panel are shown for each order.  The model line profiles do not exactly match the observed line profiles, leading to variations in the conversion values at the hydrogen lines. For use by the JWST pipeline, a smooth function was fitted to the observed conversion values for each order.}
    \label{fig:photomcal}
\end{figure}

The photometric calibration observation involved taking a short time-series observation of the star with a ramp length selected to avoid any pixel saturation.  The star is known to be non-variable at a high level of precision from the TESS satellite \citep{mullally.2022}, and the time series showed no significant variation in the signal over the roughly 5-hour period of the observation to within the standard deviation of 175\,ppm (See Sec.~\ref{sec:noiseperformance}). The extracted spectra in orders 1, 2, and 3 were obtained using a simple box aperture extraction with a box width of 40 pixels . We used a large aperture to encompass as much of the trace flux as possible in order to minimize the loss considering that the aperture correction function is ill-determined for SOSS due to the presence of multiple spectral orders. The same process was applied to a shorter observation using the GR700XD grism and the F277W filter, in order 1 only.  The output spectra in ADU/s for each of the four cases were then compared wavelength by wavelength with the stellar atmosphere model flux density in Jansky to derive the conversion values from the count rate on the detector to MJy.  These conversion values were delivered to the JWST Calibration Reference Data System (CRDS) for calibration of SOSS observations. This flux calibration gets applied in the DMS through the \texttt{photomStep} which uses the \texttt{jwst\_niriss\_photom\_nnnn.fits} reference file. The stellar model used was {\tt bd60d1753\_mod\_004.fits}, but the spectrum was smoothed to approximately the SOSS mode resolution before the conversion values were calculated.  Figure \ref{fig:photomcal} shows the extracted spectrum in instrumental units and the resulting conversion factors derived from them.

The extracted SOSS spectra were not corrected to infinite aperture because the aperture corrections cannot be derived for this mode based on the on-orbit observations.  When using the normal SUBSTRIP256 or SUBSTRIP96 sub-arrays we have no information about signal outside the sub-array area. Furthermore, in the only commissioning case where the SOSS spectra were offset to the middle of the full detector, there is significant contamination from other sources that prevents an estimate of an aperture correction.  Due to the highly distorted monochromatic PSF for this mode, it is also not possible to estimate the aperture correction from the WebbPSF simulations as can be done for the other NIRISS modes.  As a result, the photometric calibration values are specific to the extraction aperture used in the reduction of the standard star observation. Therefore, for projects requiring absolute flux calibration, we strongly encourage either 1) re-extracting the flux calibrator observation in the same manner as the science target; or 2) devising a custom aperture correction based on your science data (comparing extractions with your aperture size and that of 40 pixels used for the calibrator).

Since all SOSS mode observations put the target star at the same position on the NIRISS detector, as long as science observations use the same extraction aperture as was used for the photometric standard star---and assuming that the spectral trace of the GR700XD grism is stable over time---the photometric conversion factors should produce the correct output flux density values.  The possible small rotations of the trace from one observation to another need to be allowed for in the trace, but should not have a significant effect on the spectral extraction otherwise.  These assumptions will need to be assessed with additional calibration observations of both BD$+$60$^\circ$1753 and other standard stars in the course of observing cycle 1 and future cycles.  

\begin{figure}
    \centering
    \includegraphics[width=\linewidth]{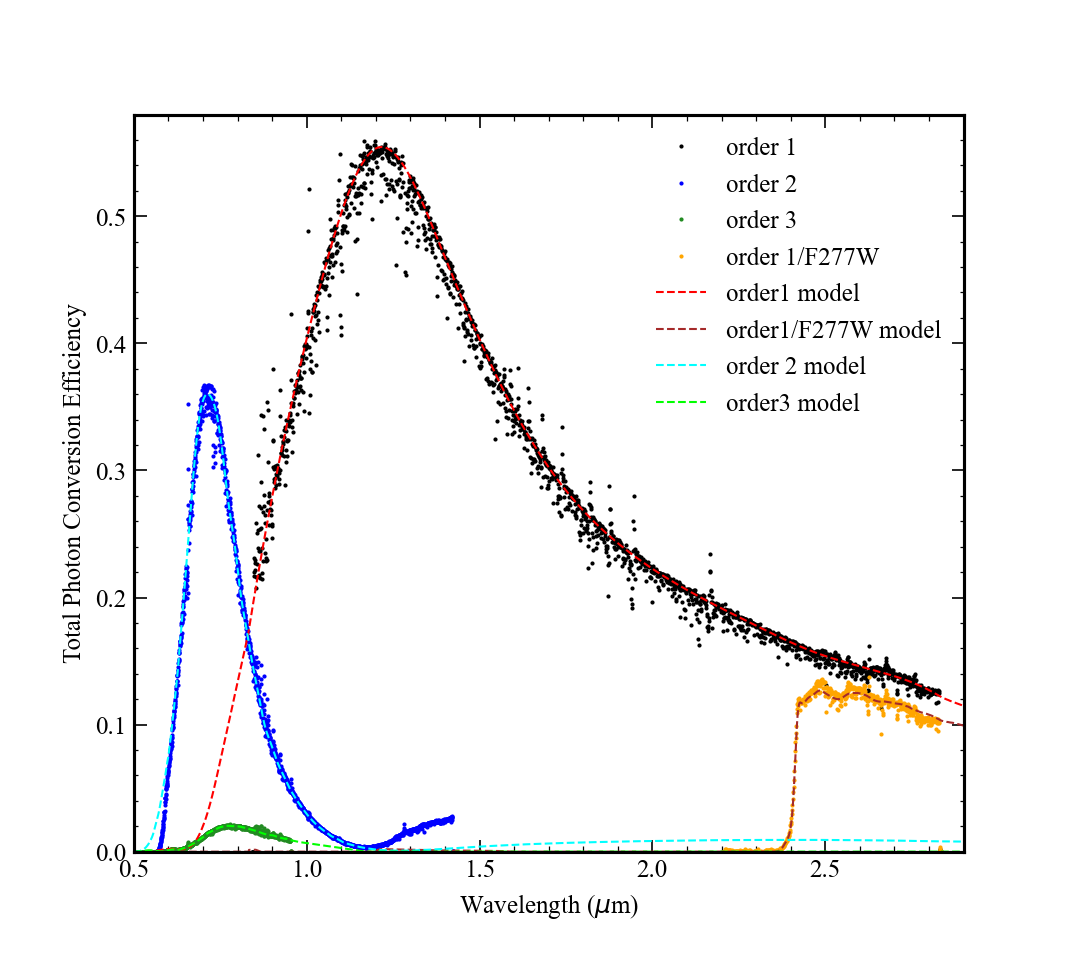}
    \caption{The estimated NIRISS total photon conversion efficiency values for the GR700XD grism, derived from the photometric calibration conversion values in Figure \ref{fig:photomcal}.}
    \label{fig:photonconversion}
\end{figure}

From the photometric conversion values, one is able to estimate the total photon conversion efficiency of the SOSS mode, defined as the number of electrons detected per photon reaching the JWST primary mirror.  Denoting the conversion value in Jy/(ADU/s) as $C(\lambda)$ and the total photon conversion efficiency as $\phi(\lambda)$ the relation between these two values is:\begin{equation}
    \phi(\lambda) = {{h\nu g}\over{10^{-26}\Delta\nu\, A\,C(\lambda)}}
\end{equation}
where $g$ is the number of electrons/ADU, $A$ is the JWST primary mirror area, and $\Delta\nu$ is the frequency interval per pixel in the spectrum.  Figure \ref{fig:photonconversion} shows the resulting values for the different orders. The transmission peaks in orders 1 and 2 come to within a few percent of the predictions based on lab measurements (See Sec~\ref{sec:opticaldesign}).

    \subsection{Spectral Trace Motion}

During a SOSS TSO, the spectral trace position remains constant well within 0.1 pixels with respect to the detector because no mechanism is moved (See Section~\ref{sec:tiltevents}). However, between science targets, the GR700XD can end at slightly different rotations due to NIRISS pupil wheel moves that occur during intervening observations, effectively causing a spectral trace rotation on the detector between visits. The finite ability of the pupil wheel to lock at the requested GR700XD position, its repeatability, was tested during Ground-based characterization. The offset was consistent with a Gaussian distribution centered around zero with a standard deviation of 0.08$\degree$ (one motor step is $\sim$0.16$\degree$). Pupil Wheel rotation mismatch translated to offsets of the spectral trace mostly along the $y$ axis by $\pm2.5$ pixels (five pixels per full motor step).

With the Fine Guidance Sensor locked in fine guiding, another potential source of motion would be if the Spacecraft Star Tracker system drifted. That would induce a field rotation centered on the Fine Guidance Sensor and perceived to first order by SOSS as an $x$ or $y$ drift of the spectral traces on the detector. Commissioning showed no evidence of this occurring.

However, an unintended behavior was observed during Commissioning when the Flight software triggered an adjustment of the GR700XD position between the  GR700XD + CLEAR science exposure and the following GR700XD + F277W  calibration exposure. Apparently, both wheels have their position fine-tuned even when a single one is commanded to move. This complicates using the GR700XD + F277W exposure to model the spectral trace for cross-order decontamination.

In-flight repeatability of the trace positioning was checked using three Commissioning observations as well as three JWST Cycle 1 TSOs. Centroids at the red end of the first order trace varied by more than 11 pixels along the spatial direction (between $y$=77 and $y$=88) among those six sequences. This is a two times larger spread than the 5\,pixels expected based on the one-motor step of Pupil Wheel repeatability. At the blue end of the trace, though, the variations were of only y=1.5\,pixels, consistent with a change in rotation of the grism. 
Similarly, a larger than expected trace motion was measured along the spectral dispersion axis where the extreme cases are offset by x=3.5\,pixels. Although the origin for these large offsets between different data sets is linked to the Pupil Wheel repeatability, the details are still under investigation. The pupil wheel rotation angle (header keyword {\tt PWCPOS}) should be compared between data sets. We emphasize that for any given data set, however, the traces are virtually locked with respect to the detector with a motion consistent with zero.

    \subsection{Wavelength Calibration} \label{sec:wavecal}
The purpose of wavelength calibration is to generate a mapping function of wavelength to detector position with sub-pixel precision to enable the science goals of SOSS. Formally, the requirement is that the wavelength solution in both orders 1 and 2 be known to be within 10\% of a resolution element. Since  NIRISS does not carry internal lamps for calibration, we turned to three astronomical sources to perform the wavelength calibration of the SOSS mode: an A~star, an M~dwarf, and an F~star.

Our main wavelength calibration target was an M~dwarf, TWA 33 (2MASS J11393382-3040002), that provided excellent feature coverage, particularly at the longer wavelengths in our bandpass, with CO, FeH and other bands, as well as reasonably strong atomic lines from species like Fe, Na, Ca, and Ti. As a safety net, we also used BD$+$60$^\circ$1753, our flux calibration standard (Sect.\ \ref{sec:fluxcal}). Even though A~stars are generally not ideal for wavelength calibration due to the paucity of absorption lines, they nevertheless do harbor clear hydrogen absorption lines. We supplemented those with the spectrum of HAT-P-14, our transit TSO test, a mid-F star.

In the process of extracting the spectra used for the following analysis, the monochromatic tilt gets averaged out. It has no other influence than, perhaps, causing negligible smearing of each spectroscopic line.

Typical wavelength calibration procedures entail fitting of well-separated individual lines from the source and using the known wavelengths to mark positions on the pixel grid. However, this approach is non-optimal for M-dwarfs where the spectra contain forests of molecular lines and bandheads, which are highly blended at our resolution. Instead, for the M-star, we adopted a `chunk-based' approach, where we calibrate adaptively-sized regions of the spectrum against a template. This allows us to treat molecular forests as a single feature, while also accommodating single-line fitting in the same process. We implement both cross-correlation and least-square minimization algorithms to calculate the shift between the observed and template chunks, and combined the information from all chunks to produce a final polynomial solution for each order.

Observations of the M-star program needed to be executed twice due to an issue with target acquisition in the first instance. Ultimately, both these observations gave consistent results except for a few pixel offsets induced by the placement of the star on the detector. We initially calibrated the M-dwarf spectra against a detailed 2D SOSS simulation as planned pre-commissioning. However, it was found that distortion and other subtle optical changes in flight (e.g., the effective input focal ratio) were causing differences between the CV3-based simulation and Commissioning observations that were impeding sub-pixel calibration precision. We thus switched to a simpler and broader approach, that of directly calibrating the lines and features observed in the A-type star (program 01091 - obs 2), the M-type star (program 01092 - obs 1 and 10), and the F-type star (program 01541 - obs 1), and forging a joint solution. 

\begin{figure*}
    \centering
    \includegraphics[width=\linewidth]{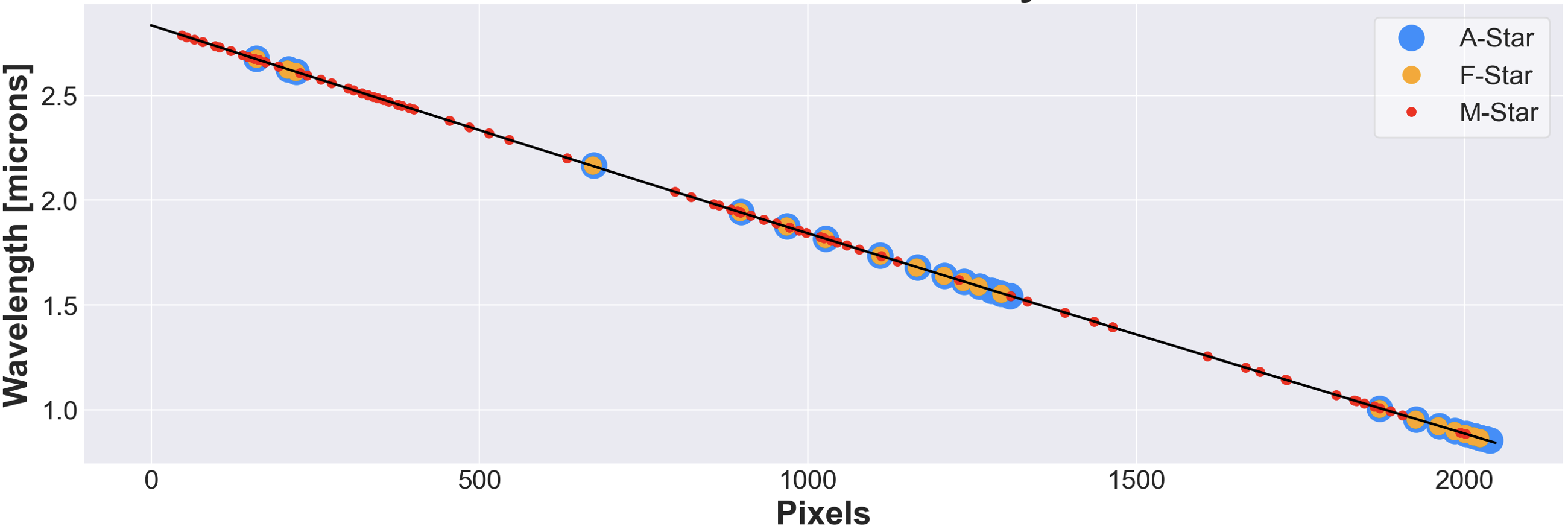}
    \caption{Example of the SOSS wavelength calibration for order 1 based on observations of three stars. Individual hydrogen absorption lines were used in the case of the A- and F-type stars while a more involved cross-correlation method was used for the M-dwarf. Here, a polynomial of second order was used to model the wavelength positions.}
    \label{fig:wavecal}
\end{figure*}

For the A and F stars, we fit the pixel positions of the hydrogen lines and assigned them to known wavelengths. For the M-dwarf we used adaptive windows to measure the features against a BT-Settl model atmosphere \citep{allard.2013} that was convolved down to the SOSS resolution and sampling. Since resolving power and dispersion are not constant across the bandpass, we treated the model atmosphere in small chunks where these quantities vary imperceptibly. The values of resolving power and dispersion were estimated for each chunk from CV3 expectations. These were close enough to in-flight conditions to predict the average behavior across tens of pixels, as opposed to the sub-pixel issues encountered before. This exercise yielded absolute wavelength versus pixel values for each of the three stars, which were then fit jointly with a low-order polynomial to independently produce wavelength calibration results for SOSS orders 1, 2, and 3 (See Fig.~\ref{fig:wavecal}). 

This wavelength solution is what is entered into the \texttt{jwst\_niriss\_spectrace\_nnnn.fits} reference table. It is then combined with the monochromatic tilt entry in the same table to produce the 2-dimensional \texttt{jwst\_niriss\_wavemap\_nnnn.fits} reference file. That later file is what gets used for the wavelength calibration through the \texttt{extract1d} DMS step.

Due to the trace motion problem discussed in the previous section, we measured and corrected for an offset of $\approx1$ pixel along the spectral axis between the 3 wavelength calibrators. That systematic error is not accounted for in the final SOSS wavelength solution which will need to be revised once a satisfying explanation for the trace motion is reached.

    \subsection{Field Contamination} \label{sec:fieldcontamination}
Because SOSS is a slitless observing mode, contamination by field sources can occur, most importantly through the zeroth spectral order of stars nearby. When such contamination overlaps one of the science target spectral traces, it causes dilution of the signal at the corresponding wavelengths which needs to be accounted for, either by subtracting a model of the zeroth order contaminant or by estimating that dilution (e.g., \cite{radica.2023}).

To roughly estimate which stars in a field of view can produce contamination, first consider that no light can enter the field of view from stars outside the pick-off mirror footprint. Second, the GR700XD deviates the target's spectral orders by roughly $+635$ pixels along the detector $y$-axis. Along the $x$-axis, the zeroth order is deviated by $\sim +800$ pixels from the target's acquisition spot. That means no field star zeroth order contamination can be projected left of the $x$ = 700 column, where the background break is indeed seen. It also means that, in FULL mode, no star below $y$\,= 535 can project its spectrum on the detector. This fact explains the fading of the background towards the bottom of the detector. No field star farther away than $\sim 250$\,pixels ($16\arcsec$) to the right of the target can contaminate because it misses the Pick Off Mirror entirely (See Fig.~\ref{fig:background}). However, stars to the left, by the full Pick Off Mirror width ($\sim 2200$ pixels - $143''$), can produce contamination from spectral orders 1 through 3. To summarize, field contamination by orders 1 to 3 occurs for field stars within a box with bounds of $-6.2~(-16.6)\,\arcsec \leq y_{\rm  target} \leq +6.2~(+16.6)\,\arcsec$ for SUBSTRIP96 (SUBSTRIP256), and $-143\,\arcsec \leq x_{\rm target} \leq +16\,\arcsec$, centered around the science target position. Zeroth order contamination occurs within a box of the same $y$-axis size but narrower along the $x$-axis: $-82\,\arcsec \leq x_{\rm target} \leq +16\,\arcsec$.

\begin{figure}
    \centering
    \includegraphics[width=\linewidth]{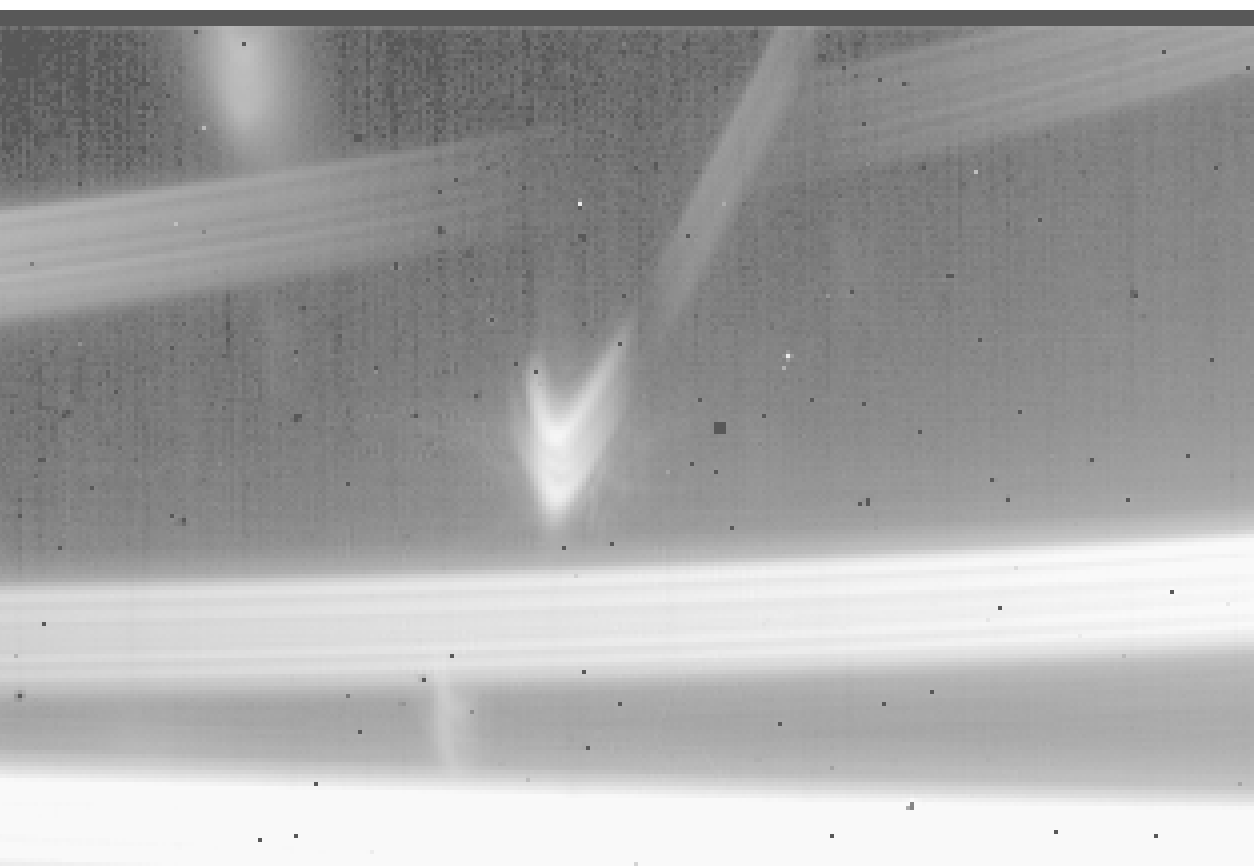}
    \caption{Example of bright zeroth order contamination on a HAT-P-1b TSO, displayed in log scale.}
    \label{fig:order0}
\end{figure}

Figure~\ref{fig:order0} shows an example zeroth order trace for a bright, isolated field star found in a HAT-P-1b TSO. Order zero gets dispersed mostly along the spatial direction ($y$-axis) because of the ZnS cross dispersing prism. The grism model and laboratory measurements (See Figure~\ref{fig:gr700xd}) reveal that the trace contains flux mostly from $\lambda \geq 2.0\,\mu$m since most of the blueward flux is transmitted in the orders 1 to 3. Nevertheless, some blue light forms a tail going up and right from the core. A minimum with almost no light appears, consistent with $\lambda = 1.2\,\mu$m.

    \subsection{Zodiacal Light Background} \label{sec:zodibackground}
\begin{figure}
    \centering
    \includegraphics[width=\linewidth]{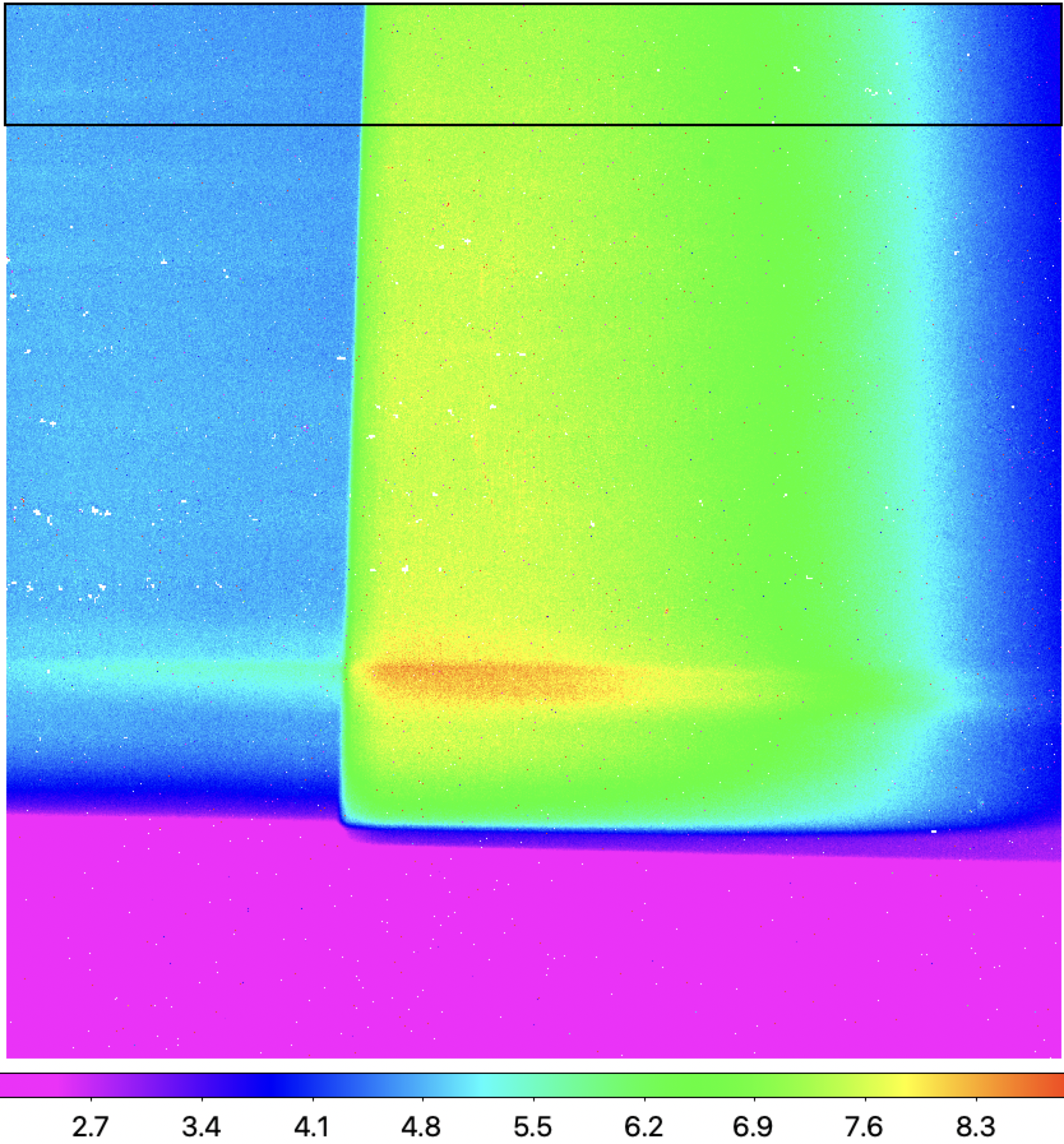}
    \caption{Background structure, in e-/s, constructed from a dithered sequence obtained at Commissioning in the FULL subarray. For reference, the SUBSTRIP256 subarray at the top is shown as a black rectangle. A break occurs at $x\approx700$ between two roughly flat plateaus. The higher level on the right side of the image is caused by zodiacal light emission in the zeroth spectral order and the break occurs when the position of that uniform background source falls off the pick-off mirror and out of the field of view. The purple region at the bottom occurs because no light can be projected by the GR700XD into that region. Note the horizontal band slightly below the center spread across the full width of the detector. It likely is the lightsaber seen in Imaging mode, i.e., a rogue path of light reaching the detector directly from a region in the sky roughly V3, V2 = (13, 3) degrees away from the telescope pointing direction\footnote{See~\url{https://jwst-docs.stsci.edu/jwst-near-infrared-imager-and-slitless-spectrograph/niriss-features-and-caveats}}.}
    \label{fig:background}
\end{figure}

The slitless nature of the SOSS mode means that zodiacal light which is reflected off the Pick-Off Mirror and passed through the GR700XD produces a column-varying background pattern. A deep (five-hour) ground-based CV3 observation in SUBSTRIP256 and a Commissioning dithered sequence obtained in program 1541 in FULL mode, both confirmed the expected shape of the structured background (See Fig.~\ref{fig:background}). It consists in two flux plateaus, on each side of a sharp, spectrally delimited, transition. The background level in the region on the left side of the subarray is about half that of the region on the right side. The transition occurs over less than two spectral pixels and forms a linear ridge, tilted by $\sim$2$^\circ$, between detector columns 705 (at the top of the detector) and 696 (at the bottom of the SUBSTRIP256 subarray). This break occurs at approximately 2.15\,$\mu$m in the first order trace and 1.09\,$\mu$m in the second order. This stepped background is naturally explained by the background light's zeroth spectral order falling off the pick-off mirror on the left side of the field of view. The plateau on the left side is almost pure first-order light while the plateau on the right is a combination of zeroth and first-order light from the background. Models reproduce this feature well. Models show that a smoothed zodiacal light spectrum shapes the background structure along the $x$-axis.

During Commissioning and from the first science programs, it was noted that the zodiacal background can change in intensity between field pointings. Simple scaling is not sufficient to track the change: separate scalings must be used on each side of the $x=700$ break. 

There is currently no step in the DMS to perform background subtraction but one is planned to be added.

\section{{\tt IDT-SOSS} Data Simulator} \label{sec:idtsoss}
To inform our data reduction pipeline and help plan Guaranteed Time Observations, we developed the Instrument Development Team SOSS simulator ({\tt IDT-SOSS}) with the goal to produce high fidelity simulations of SOSS TSOs. Simulations can be generated for all  subarrays and can be used for transit events as well as secondary eclipses. These simulations also incorporate realistic noise sources.

We use PHOENIX stellar atmosphere models generated by Peter Hauschildt (priv. comm.) based on the \cite{husser.2013} grid. They are generated at a resolving power of $R \geq 250,000$ between 0.5 and 30 $\mu$m on a grid of $3.0 \leq \log{g} \leq 5.0$ every 0.5 dex and $2300~K \leq T_{\rm eff} \leq 6900~K$ every 100~K. At each wavelength, the specific intensity is described at over $200$ angles that we fit and model using a four-parameter law \citep[Eqn. 6]{claret.2000} to reduce each model to a manageable size. The fitted specific intensity is what we use for modelling the limb darkening. We use planet atmosphere models generated at $R \geq 65,000$ with SCARLET \citep{benneke.2012,benneke.2013,benneke.2015,benneke.2019a,benneke.2019b}.

A simulation is initiated for each time step of the TSO; either each integration or each frame. We then proceed through the following steps: for each order, we first consider a padded and oversampled detector image, typically by a factor of four, and seed the expected target flux as a thin, single sub-sampled pixel, trace at the appropriate detector position. That thin trace is then convolved with monochromatic PSFs generated using \texttt{WebbPSF} \citep{perrin.2014}. We use over 100 monochromatic PSFs uniformly sampling the wavelength range of $0.5 \,\mu$m $ \leq \lambda \leq 5.5\,\mu$m. At any given wavelength, the final trace is a linear combination of the convolved traces produced at the nearest two wavelengths.

The formalism of \cite{mandel.2002} is used to model transit and eclipse light curves. The stellar atmosphere and planet atmosphere models are combined at the highest possible resolution before being seeded on the oversampled image. We generally model spectral orders 1, 2, and 3 but can also generate orders 0 and -1 --- though their exact transmission and trace positions are less well characterized.

Noiseless, as well as noisy, simulations mimicking the raw {\tt uncal.fits} JWST format are produced. The noise sources and detector artifacts considered are photon noise, readout noise, 1/$f$ noise, cosmic rays, field star contamination, zodiacal light background, quantum yield, non-linearity, dark current, detector superbias, and flat field response. We use the JWST CRDS reference files for the latter three effects and the code of \cite{rauscher.2015} for the 1/$f$ noise.

As a validation test, we simulated HAT-P-14\,b, a hot Jupiter whose time series was obtained during Commissioning (See Sec.~\ref{sec:noiseperformance}). We performed a simulation with most sources of noise except cosmic rays and zodiacal background, to simplify data reduction. Also, we opted to generate square transits (without limb darkening) to decouple the limb darkening coefficient fitting from the transit depth fitting, since these are generally correlated. For the seeded planetary spectrum, we arbitrarily chose an atmosphere with no clouds, solar metallicity, C/O = 0.3, and a temperature of 860\,K. The adopted temperature-pressure profile is that of \citet{guillot.1999} with an internal temperature of 75\,K, a heat redistribution factor of 0.25, and a Bond albedo of 0.1.

\begin{figure}
    \centering
    \includegraphics[width=\linewidth]{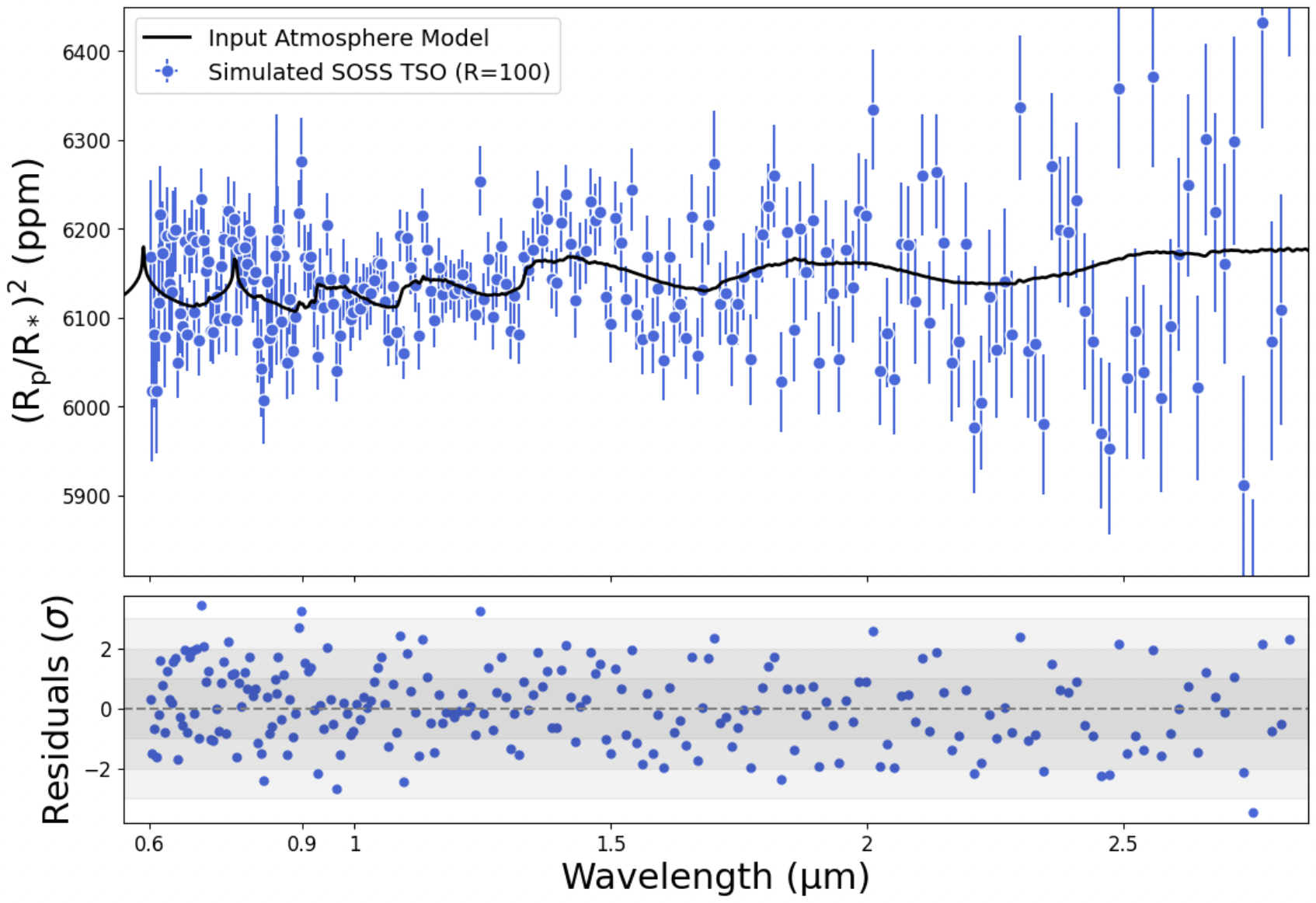}
    \caption{Top, HAT-P-14b transit spectrum, binned to a resolution of R = 100, extracted from an {\tt IDT-SOSS} simulation (blue points) compared with the seeded atmosphere spectrum in black. Bottom, residuals after subtracting the transit spectrum.}
    \label{fig:hatp14b_simu}
\end{figure}

We reduced the simulated observations using the \texttt{supreme-SPOON} pipeline\footnote{https://github.com/radicamc/supreme-spoon} (\citealp{feinstein_w39b}, Radica et al.~under revision, \citealp{coulombe.2023}) following the exact reduction steps presented in Radica et al.~(submitted). In short, we subtract the zodiacal light background, correct for 1/f noise at the group level then perform superbias subtraction, saturation flagging, jump detection, and ramp fitting. We do not apply dark subtraction because the reference file contains residual 1/f noise. We then apply flat field correction and perform bad pixel interpolation. The spectrum is extracted with the {\tt ATOCA} algorithm \citep{darveau-bernier_atoca} to explicitly model the order overlap\footnote{In the DMS, \texttt{ATOCA} is what runs for SOSS by default with the \texttt{extract1d} step. It relies on three CRDS reference files: the spectral trace profile (\texttt{jwst\_niriss\_specprofile\_nnnn.fits}), the wavelength solution (\texttt{jwst\_niriss\_wavemap\_nnnn.fits}) and the instrument throughput (\texttt{jwst\_niriss\_spectrace\_nnnn.fits}).}. An extraction box width of 35 pixels was used to match that of the Commissioning observations on the same target, See Sec.~\ref{sec:hatp14b}. The spatial profile models necessary for {\tt ATOCA} were produced with the {\tt APPLESOSS} code\footnote{https://github.com/radicamc/applesoss} \citep{radica_applesoss}. We constructed white light curves for each order by summing the flux at all extracted wavelengths for order 1, and for wavelengths between 0.6 and 0.85\,µm for order 2. We fit the white light curves using the \texttt{Juliet} code \citep{espinoza.2019}, fixing the orbital period to the simulation input value, 4.62767\,d, and the limb darkening coefficients to zero. Wide uninformative priors were used for all other parameters. We then fit the spectrophotometric light curves at the pixel level (that is, one light curve per detector column) fixing the orbital parameters to the best-fit values from the white light curve fits. The resulting transmission spectrum is shown, binned to a resolution of R\,=\,100, in Figure~\ref{fig:hatp14b_simu}, along with the atmosphere model which was seeded in the simulation.     

This simulation and transit light curve experiment demonstrate that our analysis tools can correctly retrieve the seeded planetary spectrum. In the specific case of HAT-P-14\,b, the retrieved spectrum is used as a prediction of the amplitude of atmospheric features ($\leq$\,50\,ppm) and the level of noise as a function of wavelength to be expected in the Commissioning data set.

\section{Time-Series Performance} \label{sec:noiseperformance}

    \subsection{Time Series of an A-Type Star}
During Commissioning, a five-hour time series observation of the flux standard BD$+$60$^\circ$1753, an A0VmA1V star, was obtained in the SUBSTRIP256 subarray to assess the SOSS stability. With NGROUP=3, the signal reached $\approx$\,34,000\,e$^-$ ($\approx$50\% of full well depth) above the superbias level in the brightest pixels.

To study the white light flux stability, we compared two spectral extraction methods. The first is an innovative trace-fitting approach which is described in the next section (Sec.~\ref{sec:tiltevents}) and the second is a standard box extraction method which is briefly described here.

\begin{figure*}[h]
    \centering
    \includegraphics[width=\linewidth]{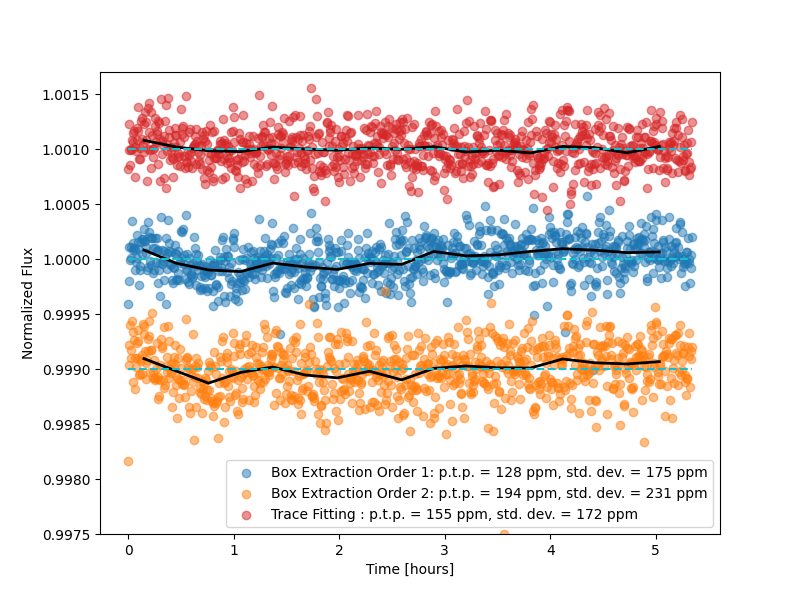}
    \caption{White light trend of the five-hour time-series of the A-type star, BD$+$60$^\circ$1753. Two spectrum extraction methods are shown. Method 1 (red) fits all orders simultaneously by applying a scale, rotation, $x$,$y$ offsets and morphology changes to a reference trace image. The second method uses a standard box extraction to extract orders 1 (blue) and 2 (orange). Order 3 is too noisy to be plotted here. Overall, for both methods, the total flux appears constant with robust standard deviations of 175~ppm and 172~ppm, respectively. While the box extraction method suggests a small downward trend of $\sim$250\,ppm for the first hour, the second method is flatter with a weaker evidence ($\sim$120\,ppm). We favor an astrophysical (stellar pulsation modes) rather than instrumental origin for this trend based on the different signal in the Fourier plane for two TSO sequences.}
    \label{fig:a0whitelight}
\end{figure*}

The data were reduced through the stage 1 DMS pipeline. We skipped the reference pixel correction and replaced it with a custom 1/$f$ noise correction applied at the group level immediately after the \texttt{saturationstep} (See Section \ref{sec:1overfcorrection}). We then applied the \texttt{flatfieldstep}, ran a custom outlier rejection based on the algorithm of \cite{nikolov.2014} and subtracted a background level by scaling the reference background map constructed during Commissioning on each side of the background break. Finally, we interpolated bad pixels. For the extraction, we used a trace width of 25\,pixels yielding optimal SNR to extract the signal for all three diffraction orders. Note that the point-to-point scatter of the out-of-transit white light curve improves fast with increasing extraction box size until it becomes almost flat for aperture sizes of $\geq25$ pixels. The exact optimal aperture size is different for each specific SOSS data set but is generally found between 25 and 35 pixels in total width. Then, for each order, we summed all extracted columns to produce a white light curve as a function of time to look for systematic trends and settling periods (See Fig~\ref{fig:a0whitelight}). 

To estimate the achieved white light noise level, we used two quantities: the standard deviation and the point-to-point (PTP) scatter. The latter simply is the standard deviation of the difference between the flux array and itself, with its indices moved by one, divided by $\sqrt{2}$. It effectively removes any systematic trend from the measurement, if any exists. It is therefore a good diagnostic tool to check for the presence of systematic deviations from the mean.

\begin{figure*}
    \centering
    \includegraphics[width=\linewidth]{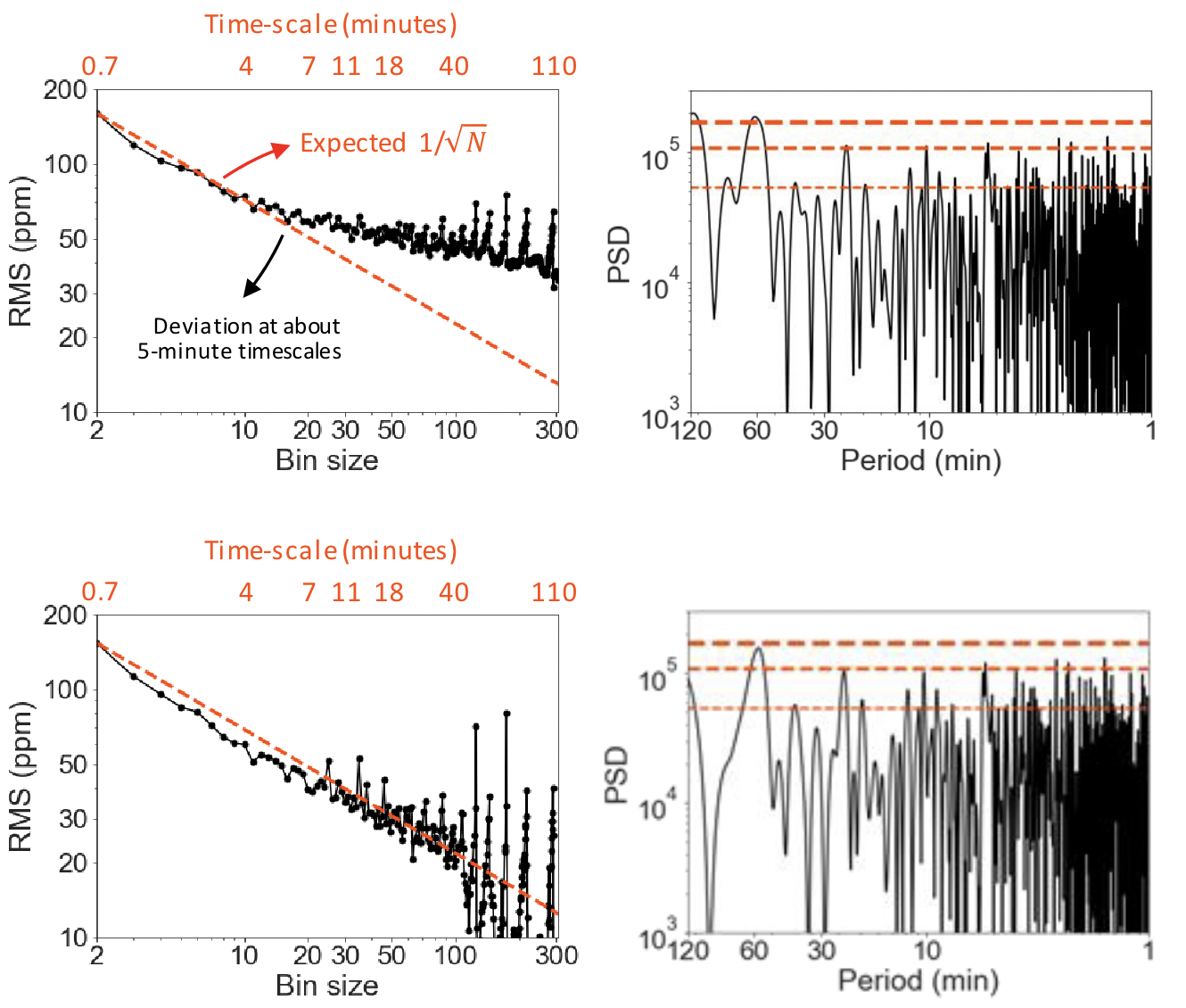}
    \caption{Noise binning of the A-type star TSO showing that SOSS achieves at least as good as 20\,ppm in systematics-free noise scatter. The top row is for the raw white light curve while the bottom row is for the white light curve after subtraction of a second-order polynomial. On the left of each row is the noise scatter as the white light curve is binned down. Noise should decrease linearly in this log-log plot if it follows photon-noise statistics. It does when the long term trend is subtracted. On the right is the power spectrum of the white light curve showing that the 60-minute component observed in the raw white light curve decreases in the trend-corrected curve.}
    \label{fig:noisebinning}
\end{figure*}

The flux is stable to within a scatter of about $175$\,ppm over the five-hour observation, consistent with the stability that we measured on TESS observations ($\leq10^{-3}$) for vetting the target. Hints of a downward trend exist during the first hour. The PTP scatter is smaller (PTP = 128\,ppm) than the standard deviation, indicating that indeed the white light curve is not entirely free of systematic trends. That trend stays consistent between the two extraction methods and between the two spectral orders for the box extraction. Whether this is a settling time of instrumental origin or a real astrophysical signal remains to be determined. We favor an astrophysical origin on the grounds of the power spectrum of this data set and that of HAT-P-14b. 
The power spectra for these two sources show power at a few dominant frequencies, perhaps stellar pulsations modes, but they are different from one another, which would probably not be the case if they were of instrumental origin. We note that ground testing was not able to see a similar settling (See Fig.~\ref{fig:stability}); however, artificial light sources are seldom of sufficient stability to allow detection at this level of precision. Expanding this careful analysis to a larger number of TSO sequences should settle the question. Fig.~\ref{fig:noisebinning} is a noise binning analysis of our white light curve. It plots the noise scatter as data is binned in time samples and it shows the power spectrum of the white light curve at full time resolution. The raw white light curve does hit a plateau at 50\,ppm. But if the long-term trend is fitted using a second-order polynomial, that curve bins down exactly as what is expected if photon-noise limited, down to 20\,ppm on a timescale of 40 minutes.

\begin{figure*}
    \centering
    \includegraphics[width=\linewidth]{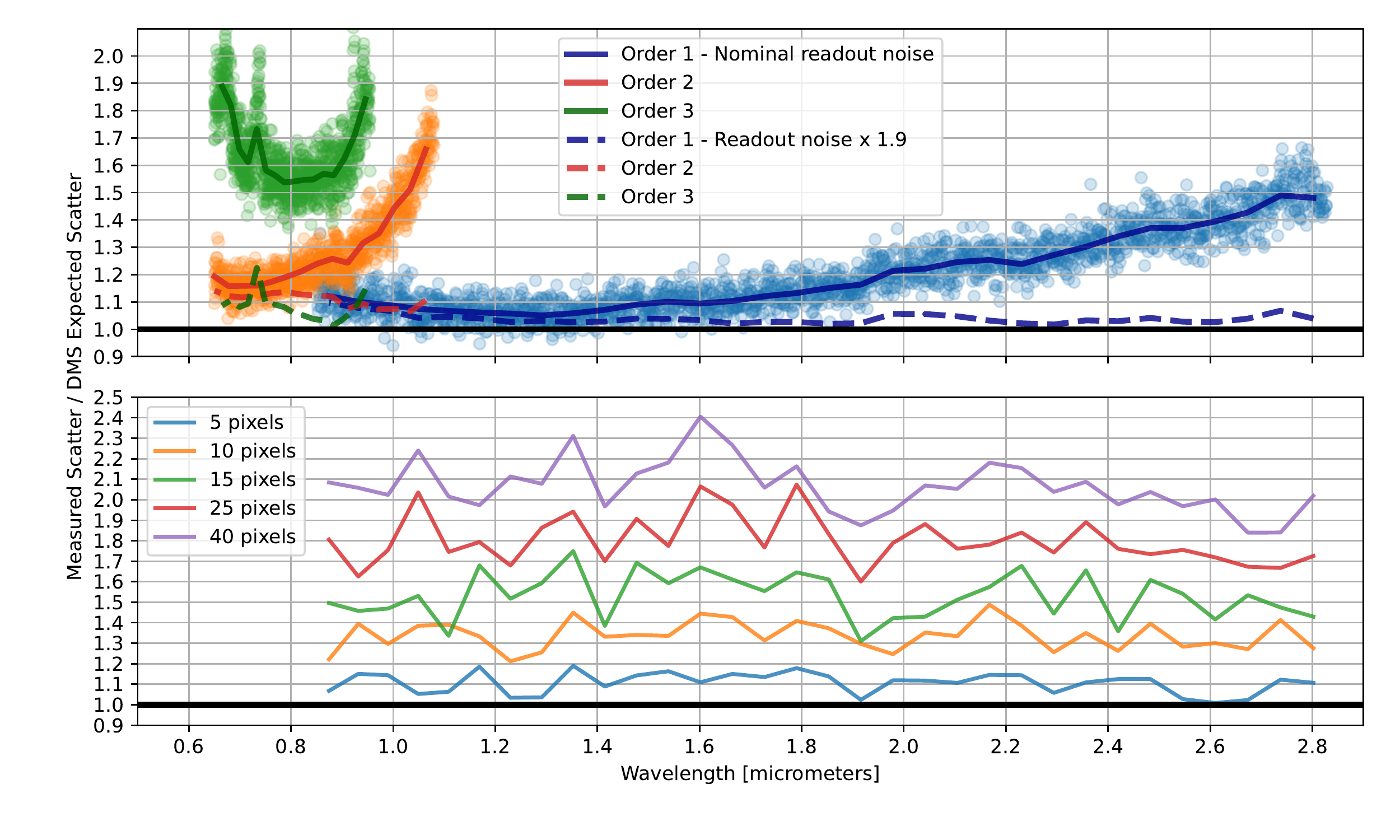}
    \caption{(Top) Noise performance on the A-type star TSO as a function of wavelength for the three spectral orders. We use as a metric the measured scatter along each spectral column divided by the spectrum uncertainties given by the DMS pipeline. The DMS only considers the photon noise and the readout noise. The actual noise is 1.05$\times$ the expected noise near the peak of the blaze function in order 1 and rises steadily to $\sim1.5\times$ at 2.8$\mu$m. For orders 2 and 3, the best ratio also coincides with the blaze wavelengths where photon noise dominates. Excess noise increases where photon noise decreases, so it behaves like a constant white noise source across the detector akin to read out noise. We attribute it to 1/$f$ correction residuals. As an experiment, if we artificially increase the expected DMS scatter by multiplying the readout noise of the CRDS reference file by 1.9$\times$ then the measured vs. expected ratio drops below 1.05$\times$ (dashed line) for wavelengths larger than 1.2~$\mu$m. The additional noise blueward may be the Fano noise associated with the quantum yield whereby more than one photo-electron is produced for each photon at bluer wavelengths. (Bottom) Noise performance on a dark exposure showing how the excess noise increases as the extraction aperture is varied between 5 to 40 pixels. The excess noise comes from high-frequency 1/f residuals that produce correlated deviations across pixels of the same column. For reference, the A0 was extracted with an aperture of 25 pixels which, on darks, results in an excess of roughly 1.9X, similar to what is found in the top panel experiment.}
    \label{fig:a0performance}
\end{figure*}

Another check of the achieved noise performance is presented in Figure~\ref{fig:a0performance}, where the scatter measured along each column of all three spectral orders is compared with the uncertainty returned by the DMS for each extracted column. A ratio of unity would suggest that all noise sources are accounted for by the DMS. However, the DMS might be optimistic as it considers only two sources of noise: photon and readout noise. Fig.~\ref{fig:a0performance} shows that the actual scatter comes close to prediction, $1.05\times$, near the blaze wavelength of order 1 where photon noise dominates, but rises to $\sim$1.5$\times$ at the red end of the trace (2.8\,$\mu$m), suggesting the presence of additional noise. Since it affects more strongly low-flux pixels, that noise is likely in the form of additional white noise, or at least, noise that affects all pixels equally, rather than a flux scaling noise. The second and third-order curves have similar shapes, albeit, compared to the expected noise, they peak at levels of $\sim$1.18$\times$ and $\sim$1.55$\times$, respectively.

These extractions were obtained from 1/$f$ noise-corrected images; a correction performed by subtracting a constant value for each detector column. It may well be that the residual unaccounted for noise is due in part to the residuals from that correction: the higher frequency 1/$f$ structures remaining in the images. If we simply assume that the noise can be modeled as an additional white noise, we find that a readout noise 1.9$\times$ larger than the CRDS readout calibration is able to reconcile the measured and expected scatters to within $\sim$5\% between $1.2\mu$m $\leq \lambda \leq 2.8\mu$m (the solid lines in Fig.~\ref{fig:a0performance}). At blue wavelengths, however, there remains a higher noise level than expected (1.05--1.2$\times$). All three spectral orders display similar trends, suggesting that the additional noise is decoupled from the pixel positions. This may be evidence of Fano noise arising when the quantum yield is different than unity. At wavelengths shorter than $\sim$1.5\,$\mu$m, it is predicted that an incident photon will produce more than one photo-electron \citep{mccullough.2008}.

To be convinced that residual 1/f noise caused the excess noise seen between $1.2\mu$m $\leq \lambda \leq 2.8\mu$m, we created a pseudo TSO of Commissioning darks containing 15 integrations and 3 groups (as the A-type star). We processed it exactly like the A-star but only varied the extraction aperture when extracting the spectrum. We then calculated the ratio of measured versus expected noise as we did for the A-type TSO. The results are shown in Fig.~\ref{fig:a0performance} - bottom panel. Clearly, as we increase the box aperture, correlated noise adds constructively and produces larger noise excess. Quantitatively, the aperture at 25 pixels produces about 1.9X the expected noise, in agreement with our experiment on the A-star where the noise reference file was artificially scaled by 1.9X to reproduce a flat curve.

We conclude that residual 1/f noise at higher temporal frequencies (small spatial structures) causes a significant noise excess for the typical box aperture of 25 pixels. To correct 1/f at all frequencies, we propose as a strategy for future TSO observations to use the FULL subarray when possible in order to use the 4-amplifier readout to construct and subtract a better model of the 1/f noise.

    \subsection{Tilt Events} \label{sec:tiltevents}
Commissioning saw several ``tilt events'' during which one of the JWST segments suddenly moved to produce, to first order, a tip-tilt misalignment \citep{rigby.2022}. It turns out that the SOSS mode is a good wavefront sensor and can detect such tilt events. The defocused PSF of the SOSS mode provides resolved information on the pupil of the instrument, and very small changes in the wavefront are expected to propagate as morphological changes in the PSF. While spatial-direction changes are not explicitly used in transit spectroscopy as this information is lost when constructing the spectrum, detecting eventual morphological changes is interesting as it provides a baseline against which one can detrend spectroscopic information. 

An innovative trace-fitting method was developed: the SOSS Inspired SpectroScopic Extraction ({\tt SOSSISSE}) (in prep). Before executing the method on the A-type star TSO, the data were processed through the DMS stage 1 pipeline which yielded a cube of integrations \textit{rateints}. We applied a custom 1/$f$ noise correction which consisted in subtracting the median value of each column on images whose astrophysical signal was removed by subtracting the stack of all integrations and where the trace was masked out. We then ran the DMS flat fielding step, performed a custom background subtraction using a reference sky frame scaled to out-of-trace pixels, and interpolated bad pixels to produce a final clean cube of integrations.

The philosophy of the trace analysis tool is to describe the SOSS time series with a model constructed from a high-signal-to-noise trace to which we add morphological changes as perturbations corresponding to the space-derivatives of the model. For example, if the trace moved by a tiny amount in the $y$ direction, one could express that exposure as the original trace plus its $y$ derivative times a constant that is proportional to the motion. In the case here, we described all frames as being a linear combination of the reference trace and its $x$, $y$ derivatives as well as the rotational derivative and the 2$^{\rm nd}$ derivative along the $y$ dimension (i.e., perpendicular to dispersion). The first three terms readily have a physical description (shifts in either direction and trace rotation relative to the center of the array). The 2$^{\rm nd}$  $y$ derivative corresponds to a change in the contrast of the PSF structures (e.g., the horns); for a Gaussian trace, this would correspond to a change in its full width at half maximum. 

The 2$^{\rm nd}$  $y$ derivative proved to be an excellent tracer of tilt events, far better than motion or rotation. As segments tilt, they affect the morphology of the trace, but they may only slightly affect the overall trace position. One key benefit of this method is that the fitted amplitude of the high-SNR trace coherently accounts for slight morphological changes and motions during the observation. Contrary to detrending, this is not a fit to the photometric signal but the inclusion of morphology changes in the PSF as part of the photometric measurement. The entire process allows only for achromatic changes in the amplitude of the trace. Of course, ultimately, the purpose of the SOSS mode is to deliver transit / eclipse/ phase curve spectroscopy; this is done in the residual space, after subtracting the mean achromatic variations of the trace model.

\begin{figure*}
    \centering
    \includegraphics[width=\linewidth]{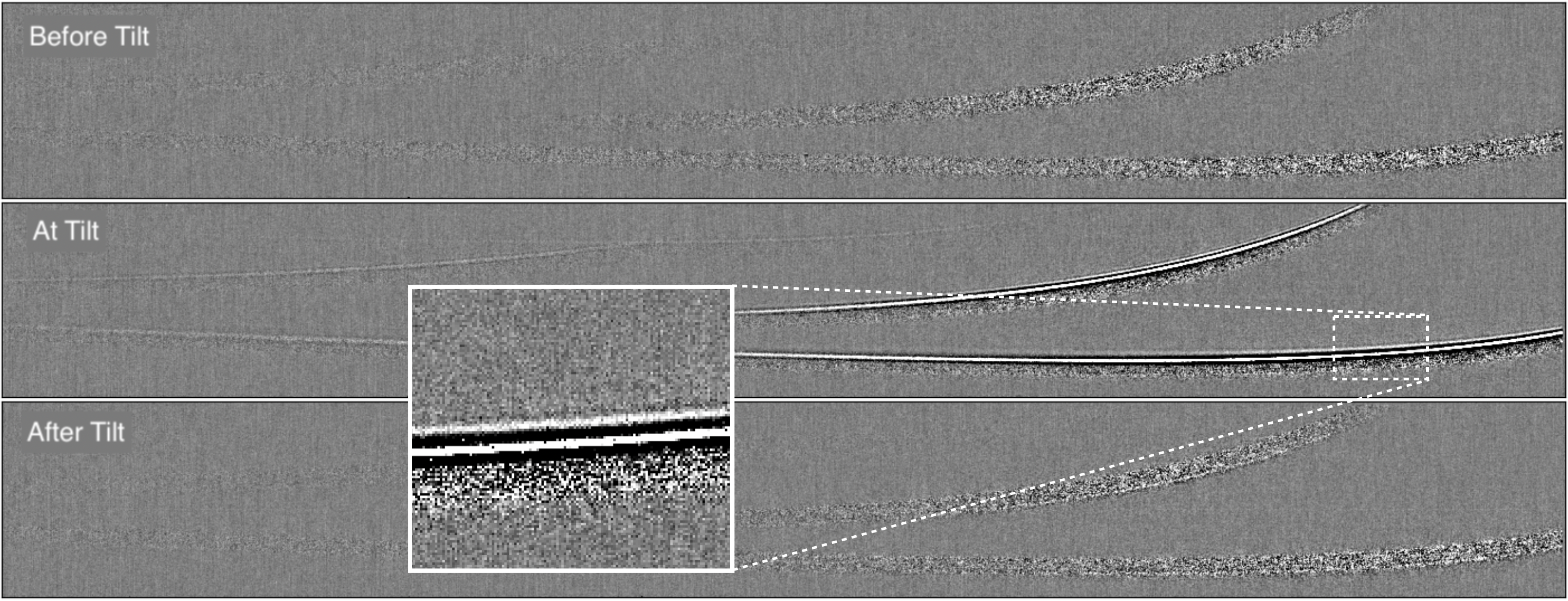}
    \caption{Difference images before, during and after the tilt event (from top to bottom). For each panel, two groups of 30 integrations are subtracted from each other after medianing each group. The trace profile clearly changed during the tilt event, for all orders and across all wavelengths. A zoom into the first order trace at 1.06\,$\mu$m is shown.}
    \label{fig:tiltdifference}
\end{figure*}

Fig.~\ref{fig:tiltdifference} illustrates how the tilt event affected the SOSS spectral traces across all spectral orders. Each panel consists of subtracting two groups of integrations, each group consisting in a median of 30 integrations. The top panel shows the difference image before the tilt event (integrations 380 to 410 minus integrations 350 to 380). It shows only white noise at the trace position, demonstrating a stable profile. The middle panel is the difference image for integrations 420 to 450 minus 380 to 410 on each side of the tilt event. It shows that one of the two trace profile horns has changed -- its core becoming darker and the immediate wings becoming brighter -- similar to a defocus term affecting only that horn presumably because the tilted mirror segment is in the upper part of the pupil. The bottom panel is the difference image after the tilt event (450 to 480 minus 420 to 450). The profile has mostly stabilized. 

\begin{figure*}
    \centering
    \includegraphics[width=\linewidth]{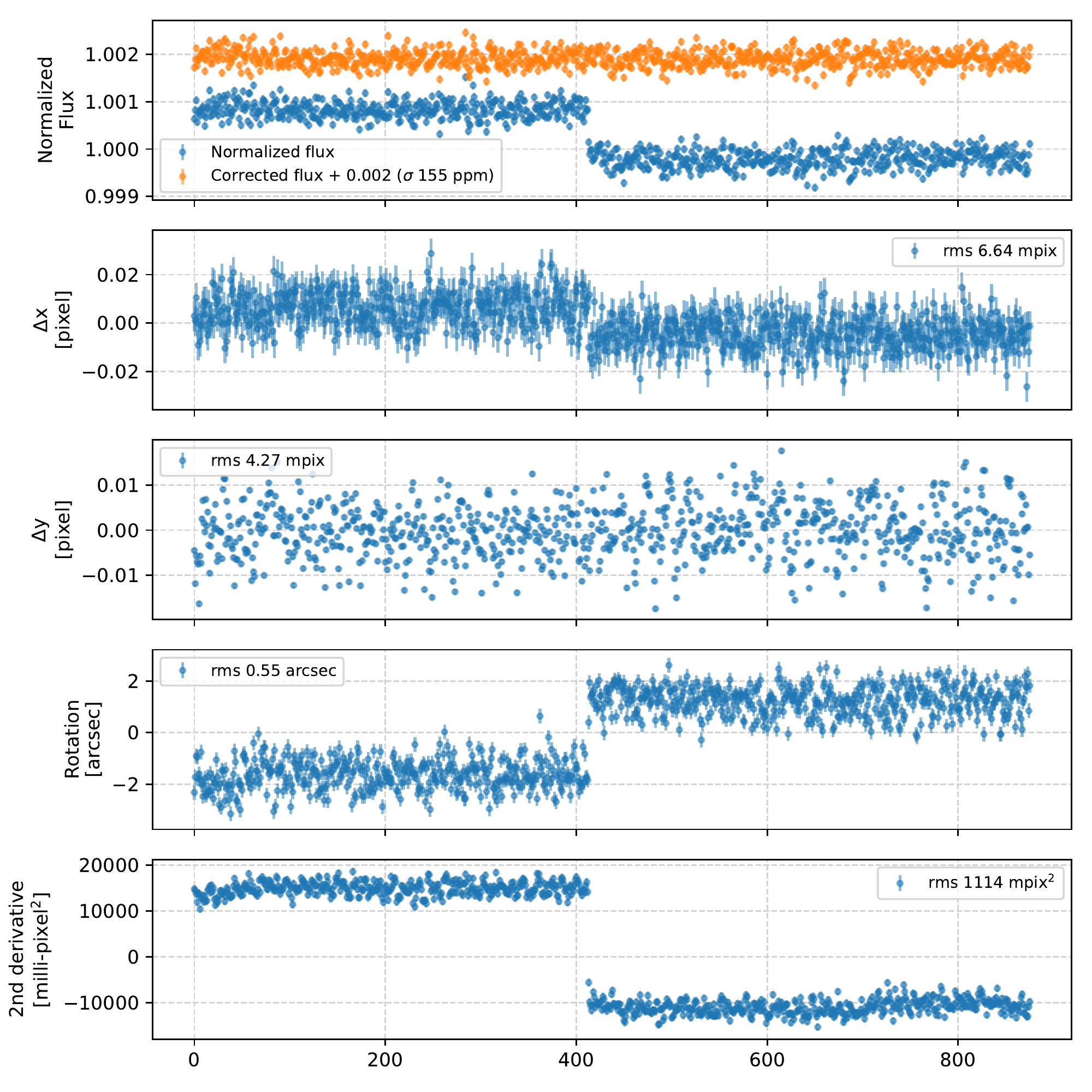}
    \caption{Times-series of the A-type star BD$+$60$^\circ$1753 where a tilt event occurred midway through the sequence. From top to bottom, the panels show the normalized white light flux, the x-axis position, the y-axis position, the trace rotation and the second derivative of the trace along the y-direction (a proxy for spatial profile shape). The tilt event near integration 413 causes an abrupt change affecting all diagnostic plots except the y position. Informed by these variations, a new flux extraction is performed to produce the corrected flux seen in orange in the top plot. A point-to-point scatter of 155\,ppm is measured. A basic standard deviation yields 172\,ppm.}
    \label{fig:a0tiltevent}
\end{figure*}

From the individual metrics presented in Fig.~\ref{fig:a0tiltevent}, one could jump to conclusions that the tilt event operated an x-position offset and a rotation of the trace profile. We argue that, instead, one of the mirror segments near the top of the pupil had its focus position shifted which changed the morphology of the trace. The individual metrics resulting from the trace fitting only adjusted to that new morphology, mimicking an x-offset $+$ rotation. We note that the spectral trace position remains extremely stable during a SOSS visit ($\sigma_x,\sigma_y = 6.6,4.3$ mas), about 1/10th of a pixel, consistent with the Fine Guidance Sensor performance.

    \subsection{HAT-P-14b Transit \label{sec:hatp14b}}
A transit of HAT-P-14b \citep{torres.2010} was observed during Commissioning (program ID: 1541) for 6.1 hours or 572 {\tt NISRAPID} integrations. It is a very dense (M=2.23M$_{Jup}$, R=1.12R$_{Jup}$; \citealp{torres.2010}) hot Jupiter and was chosen because it was observable during the Commissioning scheduling window and, most importantly, its atmospheric spectral features were expected to be small (a few tens of ppm) on account of its high density. This was to ensure that the target would not be of significant science interest, as the primary goals of these observations were to test the instrument capabilities rather than to perform a scientific analysis. Furthermore, the chosen {\tt NGROUP}=6 was selected to slightly saturate pixels near the blaze peak in order 1 in order to assess the impact of saturation and the degree of success with which saturated groups could be rejected during ramp fitting.

\begin{figure*}
    \centering
    \includegraphics[width=\linewidth]{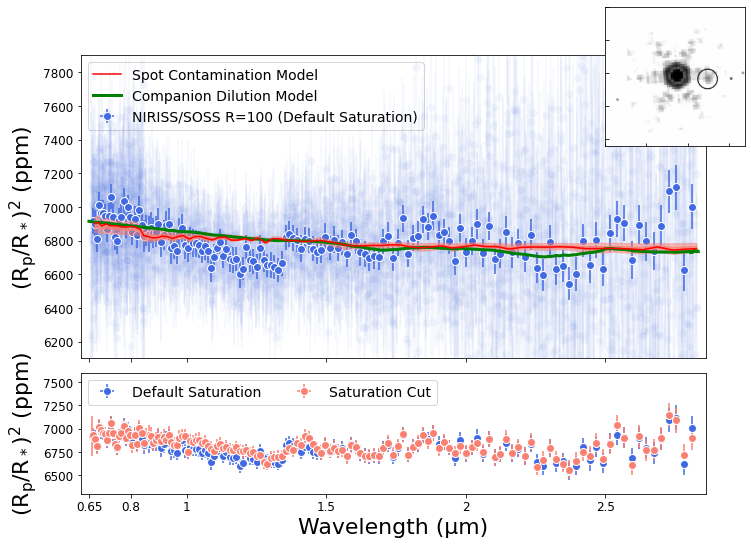}
    \caption{\emph{Top}: Transmission spectrum of HAT-P-14b from the commissioning time series at the pixel resolution (faded blue points) and binned to a resolution of R=100 (dark blue points). Overplotted in red is the best fitting spot contamination model with a spot fraction of $29^{+14}_{-13}$\% with a temperature contrast of 300\,K cooler than the star. The inset in the upper right corner shows the SOSS acquisition image. A faint co-moving companion (an early-M star; circled) is detected with a contrast of $F480M = 5.0\pm0.2$\,mag. The dilution effect means that absorption bands in the companion spectrum would translate to emission peaks in this HAT-P-14b spectrum. The estimated dilution spectrum of this companion star is shown in green.
    \emph{Bottom}: The effect of changing the threshold of saturation in the data on the final transmission spectrum. The blue spectrum uses the default DMS saturation map (saturation is flagged at roughly $\sim$45000\,ADU. In red, is the transmission spectrum resulting from a hard saturation cut a 33000\,ADU (after superbias subtraction). We find marginally deeper transits across the saturation region with this more stringent flagging.}
    \label{fig:hatp14b_MR_transitspectrum}
\end{figure*}

We once again reduced the HAT-P-14b Commissioning TSOs with the \texttt{supreme-SPOON} pipeline following the same steps as for the {\tt IDT-SOSS} simulation presented in Section~\ref{sec:idtsoss}. In order to flag saturated groups, we used the default saturation map reference file \texttt{jwst\_niriss\_saturation\_0014.fits}. We also tested implementing different saturation thresholds; for example, a more conservative hard cut of any groups with a flux level above 33,000 ADU (after superbias subtraction). However, the resulting transmission spectra are qualitatively similar (see e.g., the bottom panel of Figure~\ref{fig:hatp14b_MR_transitspectrum}). 

For the background subtraction, we noticed that when applying the \texttt{supreme-SPOON} background subtraction routine described in Radica et al.~(submitted), using the default STScI SUBSTRIP256 background model, the ``step'' in the zodiacal background was not being adequately removed. We therefore implemented a slight modification to the background subtraction procedure to separately scale both sides of the STScI model such that the background step is properly corrected. We used the region x,y=(300:500,200:240) to scale the red end, and x,y=(1000:1100,135:215) to scale the blue end of the model.

\begin{figure*}
    \centering
    \includegraphics[width=\linewidth]{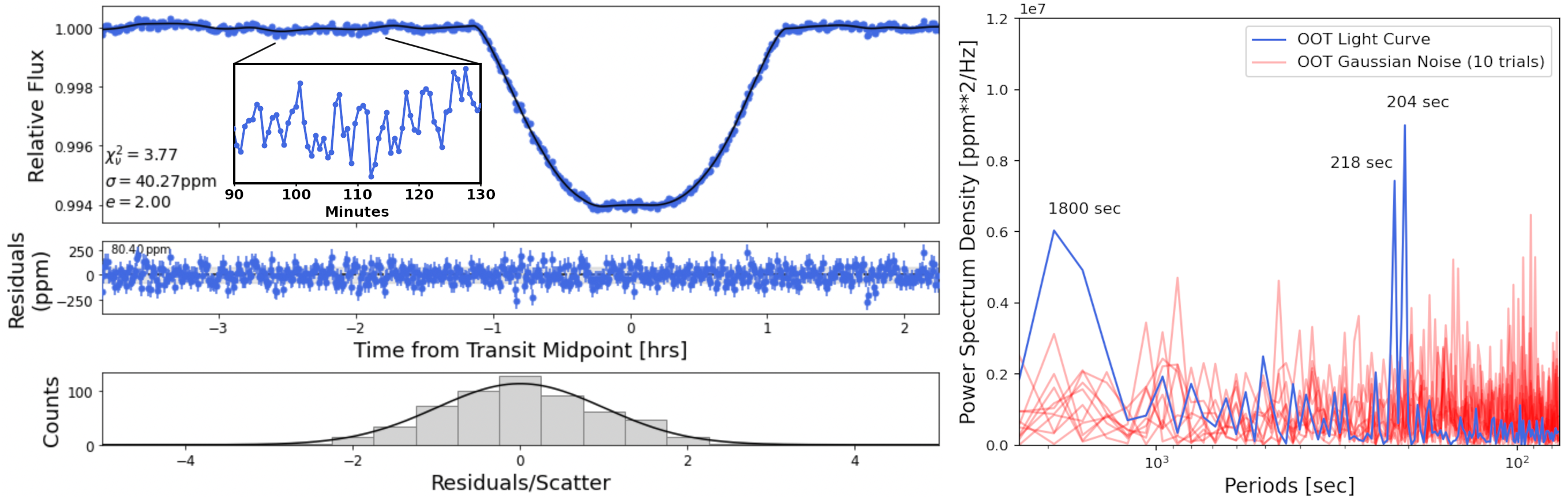}
    \caption{White light curve fitting results. \emph{Top left}: HAT-P-14b order 1 white light curve from the Commissioning TSOs. In black is the best fitting transit plus GP systematics model. The transit model itself is shown in the blue dashed line, and the GP model in the red dashed line. The final fit has a $\chi^2_\nu = 2.41$. In the panel, $\sigma$ is the median error bar size and $e$ is the error multiple necessary to reach a $\chi^2_\nu$ equal to unity. The inset shows details of correlated noise.
    \emph{Middle left}: The light curve fit residuals. They are evenly distributed about zero, with no clear trends present. 
    \emph{Bottom left}: Histogram of the residuals, showing that the systematics model has adequately captured the baseline variations. \emph{Right}: The power spectrum density of the order 1 out-of-transit white light curve (the first 2.7h) detects 3 significant peaks at 204, 218 and $\approx$ 1800 seconds periods. The origin of these peaks is unknown but the two 200-sec peaks match the oscillations seen in the top left inset.}
    \label{fig:hatp14b_MR_whitelight}
\end{figure*}

As with the simulated observations, we extracted the stellar spectra using the {\tt ATOCA} and {\tt APPLESOSS} algorithms, with an extraction width of 35 pixels. For HAT-p-14b, that was the width at which we found a minimum in the white light SNR. We also followed the same light curve fitting procedure, first constructing and fitting white light curves for each order, then fitting the spectrophotometric light curves at the pixel level, fixing the orbital parameters from the white light curve results. We again only consider wavelengths for order 2 from 0.6 -- 0.85\,µm. For the white light curve fits, we fix the orbital period to 4.62767\,d, and set wide, uninformative priors on the other parameters: mid-transit time, $T_0$, impact parameter, $b$, scaled planetary radius, $R_p/R_*$, and the scaled orbital semi-major axis, $a/R_*$. We fit two parameters of the quadratic limb-darkening law, following the parameterization of \citet{kipping_ld}, as well as a scalar jitter term which is added in quadrature to the error bars on each data point. We furthermore fit two parameters of a Matérn-3/2 Gaussian process (GP) as implemented by \citet{celerite} in order to model significant variations that we notice in the out-of-transit (OOT) baseline. In total, we fit nine parameters to each white light curve, again using the nested sampling routine \texttt{dynesty} \citep{speagle_dynesty} as implemented in \texttt{Juliet}. The order 1 white light curve and the best-fitting model are shown in Fig.~\ref{fig:hatp14b_MR_whitelight} while the posterior distribution of the free parameters are listed in Tab.~\ref{tab:freeparams}.

The white light curve shows what appears to be correlated noise (inset in the top-left of Fig.~\ref{fig:hatp14b_MR_whitelight} ). The origin of the extra noise seen in the OOT was studied by calculating the power spectrum density which shows 3 significant peaks: 2 narrow peaks at 204 and 218 seconds and a wide peak at about 1800 seconds (Fig.~\ref{fig:hatp14b_MR_whitelight} - right). The origin of those peaks is unknown. Similar 200- sec signals have been identified in a few (but not all) SOSS TSO data sets (priv. comm.)\footnote{Investigation of this extra noise is under way at STScI (Espinoza et al. in prep.)}

\begin{table}

\caption{\label{tab:freeparams} Free parameters used in the HAT-P-14b transit white light curve analysis}
\small
\begin{tabular*}{0.45\textwidth}{l|c}
\hline
\hline
Mid transit time [MJD] & $59738.4622\pm0.0002$ \\
Radius ratio \textit{(R$_{\text{p}}$/R$_*$)} & $0.081^{+0.003}_{-0.002}$ \\
Semi-major axis (\textit{a/R$_*$)} & $8.15^{+0.15}_{-0.12}$ \\
Impact parameter (b) & 0.913$^{+0.004}_{-0.004}$ \\
Limb darkening q1  & 0.063$^{+0.078}_{-0.045}$ \\
Limb darkening q2  & 0.62$^{+0.27}_{-0.40}$ \\
Jitter ($\sigma$) [ppm] & 80.4$^{+2.6}_{-2.5}$ \\
GP$_{\sigma}$ &  184$^{+55}_{-40} \times 10^{-6}$ \\
GP$_{\rho}$ & 2.15$^{+0.25}_{-0.11} \times 10^{-2}$\\
\hline
\end{tabular*}

\end{table}

We then fit the spectrophotometric light curves at the pixel level, fixing the orbital parameters to the best fitting values from the white light curve fits. For the limb-darkening, we set Gaussian priors around the values predicted by 1D-stellar models \citep{kurucz1993atlas9} using ExoTiC-LD \citep{laginja_exotic, hannah_wakeford_2022_6809899}. The width of the prior is determined by the spread in modeled coefficients resulting from varying the stellar effective temperature, gravity, and metallicity according to the spread in these values in the literature. For the GP parameters, we set Gaussian priors using the best fitting value, and 1$\sigma$ bounds from the white light curve fit as the Gaussian position and width. The resulting transmission spectrum is shown in Figure~\ref{fig:hatp14b_MR_transitspectrum} at the pixel resolution, as well as binned to R=100.

Simulations of HAT-P-14b predict shallow water features at a level of no more than 50\,ppm, and even smaller for cloudy atmospheres. However, our transmission spectrum is not such a flat, featureless spectrum. Deviations at the $\sim$100\,ppm level are noticeable, namely a rising slope towards the blue as well as a break at 1.4\,$\mu$m. The blue slope extends well below 0.8\,$\mu$m in the second order wavelength coverage. Since it is present in, and continuous across both spectral orders, this indicates that it does not arise from a detector artifact. 

At first glance, the blue slope is suggestive of unocculted star spots. We fitted a spot model based on \citet{rackham.2018} with spots cooler than the star surface and occupying a fraction of the visible surface. Qualitatively, that model only partly reproduces the blue slope. The best fit returns spots with $\Delta$T = 300\,K and occupying a fraction of $29^{+14}_{-13}$\% (overplotted in red in the top panel of Figure~\ref{fig:hatp14b_MR_transitspectrum}). But it is not very satisfactory with a reduced $\chi^2=1.98$. In addition, this model can not explain the break seen at 1.4\,$\mu$m.

Alternatively, we note that HAT-P-14 harbors a faint companion star which could dilute the transit spectrum \citep{ngo.2015}. Adaptive optics imaging shows a companion at a separation of $0.857\pm0.009$ arcseconds with a contrast of $\Delta H=5.24\pm0.09$\,mag, $\Delta K_s=5.75\pm0.05$\,mag. Indeed, this companion is seen in the $64\times64$\,pixel SOSS acquisition subarray at exactly the same position with a contrast of $\Delta F480M=5.0\pm0.2$\,mag (see inset of Figure~\ref{fig:hatp14b_MR_transitspectrum}). This confirms that the companion is bound to HAT-P-14 and has the colors of an early-M star. Coupling this 1\% dilution factor from the companion with the $\sim$1\% transit depth of HAT-P-14b can explain features appearing at the $\sim$100\,ppm level in the planet's transmission spectrum. In this case, we expect spectral features that are scaled and inverted versions of the M-type companion spectrum which harbors water bands at 1.4\,$\mu$m and 1.9\,$\mu$m. We produced an {\tt IDT-SOSS} simulation, adding a 3300\,K dwarf at the correct offset position from HAT-P-14 and extracted its spectrum along with that of our previous HAT-P-14b simulation. The ratio of the spectra was scaled by 1\% to account for the transit depth and inverted to mimic the dilution effect. That model is overplotted green in the top panel of Figure~\ref{fig:hatp14b_MR_transitspectrum}. The position and amplitude of the water bands reproduce the spectrum well and it also features a similar blue slope. However, it still struggles to reproduce the break at 1.4\,$\mu$m.

Our initial experiment to test whether mild saturation is problematic comes short from a final answer. The transit spectrum is not flat and its features are stronger and different from those expected from simulations, thus likely not intrinsic. Two contamination models can be invoked: an unocculted star spot or a field M-star contamination. They both reduce the tensions but still leave important unexplained features such as the 1.4~$\mu$m break. Treatment of the saturation during data reduction also impacts the extracted spectrum. Therefore, quantifying the effect of the saturation is difficult. Letting SOSS TSO observations saturate is therefore risky for high precision spectroscopy. It definitely complicates the analysis and likely introduces systematic noise at some level ($\approx 100$~ppm). Additional work to better handle saturation is required.

In full disclosure, the HAT-P-14b observation did produce saturation starting at NGROUP=3. Saturation first occurred in the two horns of the trace profile and progressed to affect all pixels of the profile by NGROUP=6 for all wavelengths between 0.9~$\mu$m and 1.5~$\mu$m of order 1. That wavelength range is also where the 1.4\,$\mu$m break appears. By which detector physics process is saturation affecting the measurements is uncertain. Persistence, imperfect non-linearity correction or charge migration (a.k.a. the Brighter-Fatter effect, \cite{coulton.2018} and references therein) could be at play.

\section{Conclusions} \label{sec:conclusions}
The SOSS mode of NIRISS was designed specifically to undertake time-series observations of exoplanets between $0.6~\mu$m and $2.8~\mu$m. Its optics consist of a ZnSe grism coupled with a cross-dispersing ZnS prism to project three spectral diffraction orders on the detector, in addition to a weak cylindrical surface to spread the light across roughly 23 pixels in the cross-dispersion direction, thereby enabling observations of stars as bright as J=6.7 (7.5) in SUBSTRIP96 with NGROUP=1 (NGROUP=2) at a resolving power of 650 at $\lambda = 1.2\,\mu$m. Three subarray modes are offered: SUBSTRIP96, SUBSTRIP256, and FULL. Ground testing and initial Commissioning observations demonstrate that the SOSS mode is operating at nominal performance and on par with other JWST time-series modes, such as the NIRSpec Bright Object Time-Series mode which delivers close to photon-limited noise performance \citep{espinoza.2022}. The end-to-end optics transmission produces a photon conversion efficiency $\geq$50\% at the blaze wavelength (1.2\,$\mu$m) of the first order and $\geq$35\% at 0.7\,$\mu$m in the second order. 

Excellent white light flux stability with time was demonstrated with a 5-h time-series observation on BD$+$60$^\circ$1753, a A-type star flux standard. After correcting for a weak long-term trend with a second-order polynomial, the scatter of the binned flux scales as one over the square root of the number of photons, as expected for photon statistics. This was tested down to $\approx20$~ppm on a 40-minute time scale. Whether the weak evidence for a $\approx250$~ppm downward trend during the first hour of that TSO is of instrumental origin would need to be confirmed with more observations. Another TSO, on the exoplanet host HAT-P-14b, shows evidence for extra correlated noise linked to a pair of peaks at 204 and 218 seconds in the power spectrum density plot. Its cause is under investigation but we can already say that not all TSOs are affected.

The 1/$f$ noise correction, cosmic ray flagging, bad pixel interpolation, and zodiacal background subtraction are the most important data processing steps to master for producing near photon-noise limited analyses. Even if carefully done, the 1/$f$ correction is imperfect. Approximating the 1/$f$ signal as a constant across a SOSS column leaves noise residuals underneath the spectral traces. That has the same effect as almost doubling ($\times1.9$) the read-out noise in each integration.

While the spectral trace position remains stable to within $\approx0.1$ pixel during a single SOSS visit, variations are measured between visits. These positional excursions correlate with the dialed pupil wheel rotation which varies by up to 2 step motors. That has the effect of rotating the GR700XD, producing a trace rotation with respect to the detector. The measured peak-to-peak spectral trace position variations between half a dozen TSOs are $\Delta x,\Delta y = 3.5,10$\,pixels. This complicates the wavelength calibration but, fortunately, the {\tt PWCPOS} header keyword can be used to retrieved this rotation. Until the problem is fully characterized and a fix implemented in the DMS, the wavelength calibration suffers from a systematic error of a few pixels.

It was discovered that SOSS can be a very good wavefront sensor --- it can detect JWST mirror segment tilt events through small changes in the morphology of the spectral trace profile. These tilt events can produce flux jumps in the time-series. We presented a tool to diagnose these events.

A time-series targeting the massive planet HAT-P-14b was obtained to characterize the effect of letting pixels in the peak of the blaze function saturate. Our simulations predicted a very small planetary atmospheric signal of $\leq 50 ppm$ but features at least twice that big were observed. Analysis was complicated due to contamination by a faint companion and hints of unocculted spot crossing. The exact recipe adopted for saturation mitigation also produced effects at a comparable level. Further work is required but saturation definitely complicates analysis.

Science programs based on SOSS TSOs have started. The NIRISS instrument development team allocated almost 200 hours of NIRISS guaranteed time observation to the \emph{NIRISS Exploration of the Atmospheric diversity of Transiting exoplanets} (NEAT; program ID: 1201; PI Lafrenière). Understanding whether close-in exoplanets form in situ or migrate from an outer orbit after formation is one of the key open questions in exoplanetary science. The carbon-to-oxygen ratio (C/O) has been proposed as a good proxy for the formation environment. Volatiles in the protoplanetary disk like H$_2$O, CH$_4$, NH$_3$, CO, and CO$_2$ will condense into statistically O-rich condensates beyond the ice line whereas condensates closer in will be carbon-rich. NEAT aims to address that question by measuring C/O ratios for several close-in and nearby exoplanets. NEAT will spearhead the use of NIRISS/SOSS, but will also carry some NIRSpec Prism observations. Disk-averaged, secondary eclipse observations that probe deeper into the atmospheres will complement transit observations which probe the terminator at low pressures. Phase curve observations of one hot Jupiter and one hot Neptune will also be conducted. We will target a diversity of 14 planets spanning a wide range of sizes (terrestrial to Jupiter size) and temperature (200 K to 2000 K).

\section{Acknowledgements}

This project is undertaken with the financial support of the Canadian Space Agency. M.R.~acknowledges financial support from NSERC. FRQNT, and iREx. D.J.\ is supported by NRC Canada and by an NSERC Discovery Grant. J.D.T was supported for this work by NASA through the NASA Hubble Fellowship grant \#HST-HF2-51495.001-A awarded by the Space Telescope Science Institute, which is operated by the Association of Universities for Research in Astronomy, Incorporated, under NASA contract NAS5-26555. R. A. is a Trottier Postdoctoral Fellow and acknowledges support from the Trottier Family Foundation. This work was supported in part through a grant from FRQNT.

\bibliographystyle{aasjournal}
\bibliography{bibliography, mcr.bib}

\end{document}

%% file: authors.tex
\author[0000-0003-0475-9375]{Lo\"ic Albert}
\affiliation{D\'epartement de Physique and Observatoire du Mont-M\'egantic, Universit\'e de Montr\'eal, C.P. 6128, Succ. Centre-ville, Montr\'eal, H3C 3J7, Québec, Canada.}
\affiliation{Institut Trottier de Recherche sur les exoplan\`etes, Universit\'e de Montr\'eal}

\author[0000-0002-6780-4252]{David Lafreni\`ere}
\affiliation{D\'epartement de Physique and Observatoire du Mont-M\'egantic, Universit\'e de Montr\'eal, C.P. 6128, Succ. Centre-ville, Montr\'eal, H3C 3J7, Québec, Canada.}
\affiliation{Institut Trottier de Recherche sur les exoplan\`etes, Universit\'e de Montr\'eal}

\author[0000-0001-5485-4675]{Doyon, Ren\'e }
\affiliation{D\'epartement de Physique and Observatoire du Mont-M\'egantic, Universit\'e de Montr\'eal, C.P. 6128, Succ. Centre-ville, Montr\'eal, H3C 3J7, Québec, Canada.}
\affiliation{Institut Trottier de Recherche sur les exoplan\`etes, Universit\'e de Montr\'eal}

\author[0000-0003-3506-5667]{\'Etienne Artigau}
\affiliation{D\'epartement de Physique and Observatoire du Mont-M\'egantic, Universit\'e de Montr\'eal, C.P. 6128, Succ. Centre-ville, Montr\'eal, H3C 3J7, Québec, Canada.}
\affiliation{Institut Trottier de Recherche sur les exoplan\`etes, Universit\'e de Montr\'eal}

\author[0000-0002-3824-8832]{Kevin Volk}
\affil{Space Telescope Science Institute, 3700 San Martin Drive, Baltimore, MD 21218, USA}

\author[0000-0002-5728-1427]{Paul Goudfrooij}
\affil{Space Telescope Science Institute, 3700 San Martin Drive, Baltimore, MD 21218, USA}

\author{Andr\'e R. Martel}
\affiliation{Space Telescope Science Institute, 3700 San Martin Drive, Baltimore, MD 21218, USA}

\author[0000-0002-3328-1203]{Michael Radica}
\affiliation{D\'epartement de Physique and Observatoire du Mont-M\'egantic, Universit\'e de Montr\'eal, C.P. 6128, Succ. Centre-ville, Montr\'eal, H3C 3J7, Québec, Canada.}
\affiliation{Institut Trottier de Recherche sur les exoplan\`etes, Universit\'e de Montr\'eal}

\author[0000-0002-5904-1865]{Jason Rowe}
\affiliation{Department of Physics \& Astronomy, Bishop's University, Sherbrooke, QC J1M 1Z7, Canada.}

\author[0000-0001-9513-1449]{N\'estor Espinoza}
\affil{Space Telescope Science Institute, 3700 San Martin Drive, Baltimore, MD 21218, USA}

\author[0000-0001-8127-5775]{Arpita Roy}
\affil{Space Telescope Science Institute, 3700 San Martin Drive, Baltimore, MD 21218, USA}
\affil{Department of Physics and Astronomy, Johns Hopkins University, 3400 N Charles St, Baltimore, MD 21218, USA}

\author[0000-0002-0201-8306]{Joseph C. Filippazzo}
\affil{Space Telescope Science Institute, 3700 San Martin Drive, Baltimore, MD 21218, USA}

\author[0000-0002-7786-0661]{Antoine Darveau-Bernier}
\affiliation{D\'epartement de Physique and Observatoire du Mont-M\'egantic, Universit\'e de Montr\'eal, C.P. 6128, Succ. Centre-ville, Montr\'eal, H3C 3J7, Québec, Canada.}
\affiliation{Institut Trottier de Recherche sur les exoplan\`etes, Universit\'e de Montr\'eal}

\author[0000-0003-4787-2335]{Geert Jan Talens}
\affil{Department of Astrophysical Sciences, Princeton University, 4 Ivy Lane, Princeton, NJ 08544, USA}

\author[0000-0003-1251-4124]{Anand Sivaramakrishnan}
\affiliation{Space Telescope Science Institute, 3700 San Martin Drive, Baltimore, MD 21218, USA}
\affiliation{Astrophysics Department, American Museum of Natural History, 79th Street at Central Park West, New York, NY 10024}
\affil{Department of Physics and Astronomy, Johns Hopkins University, 3400 N Charles St, Baltimore, MD 21218, USA}

\author[0000-0002-4201-7367]{Chris J. Willott}
\affiliation{NRC Herzberg Astronomy and Astrophysics, 5071 West Saanich Rd, Victoria, BC, V9E 2E7, Canada}

\author[0000-0003-2429-7964]{Alexander W. Fullerton}
\affiliation{Space Telescope Science Institute, 3700 San Martin Drive, Baltimore, MD 21218, USA}

\author[0000-0002-5907-3330]{Stephanie LaMassa}
\affil{Space Telescope Science Institute, 3700 San Martin Drive, Baltimore, MD 21218, USA}

\author{John B. Hutchings}
\affiliation{NRC Herzberg Astronomy and Astrophysics, 5071 West Saanich Rd, Victoria, BC, V9E 2E7, Canada}

\author[0000-0002-1715-7069]{Neil Rowlands}
\affiliation{Honeywell Aerospace \#100, 303 Terry Fox Drive, Ottawa,  ON  K2K 3J1, Canada} 

\author[0000-0003-3504-1569]{M. Bego\~na Vila}
\affiliation{NASA Goddard Space Flight Center, 8800 Greenbelt Rd, Greenbelt, MD 20771}
\affiliation{KBR Space Engineering Division, 8120 Maple Lawn Blvd, Fulton, MD 20759}

\author{Julia Zhou}
\affiliation{Honeywell Aerospace \#100, 303 Terry Fox Drive, Ottawa,  ON  K2K 3J1, Canada} 

\author{David Aldridge}
\affiliation{Honeywell Aerospace \#100, 303 Terry Fox Drive, Ottawa,  ON  K2K 3J1, Canada} 

\author{Michael Maszkiewicz}
\affiliation{Canadian Space Agency, 6767 Route de l'Aéroport, Saint-Hubert, QC J3Y 8Y9, Canada}

\author{Mathilde Beaulieu}
\affiliation{Université Côte d'Azur, Observatoire de la Côte d'Azur, CNRS, Laboratoire Lagrange, F-06108 Nice, France.}

\author[0000-0003-4166-4121]{Neil J. Cook} 
\affiliation{D\'epartement de Physique and Observatoire du Mont-M\'egantic, Universit\'e de Montr\'eal, C.P. 6128, Succ. Centre-ville, Montr\'eal, H3C 3J7, Québec, Canada.}
\affiliation{Institut Trottier de Recherche sur les exoplan\`etes, Universit\'e de Montr\'eal}

\author[0000-0002-2875-917X]{Caroline Piaulet}
\affiliation{D\'epartement de Physique and Observatoire du Mont-M\'egantic, Universit\'e de Montr\'eal, C.P. 6128, Succ. Centre-ville, Montr\'eal, H3C 3J7, Québec, Canada.}
\affiliation{Institut Trottier de Recherche sur les exoplan\`etes, Universit\'e de Montr\'eal}

\author[0000-0001-6809-3520]{Pierre-Alexis Roy} 
\affiliation{D\'epartement de Physique and Observatoire du Mont-M\'egantic, Universit\'e de Montr\'eal, C.P. 6128, Succ. Centre-ville, Montr\'eal, H3C 3J7, Québec, Canada.}
\affiliation{Institut Trottier de Recherche sur les exoplan\`etes, Universit\'e de Montr\'eal}

\author{Pierrot Lamontagne} 
\affiliation{D\'epartement de Physique and Observatoire du Mont-M\'egantic, Universit\'e de Montr\'eal, C.P. 6128, Succ. Centre-ville, Montr\'eal, H3C 3J7, Québec, Canada.}
\affiliation{Institut Trottier de Recherche sur les exoplan\`etes, Universit\'e de Montr\'eal}

\author{Kim Morel} 
\affiliation{D\'epartement de Physique and Observatoire du Mont-M\'egantic, Universit\'e de Montr\'eal, C.P. 6128, Succ. Centre-ville, Montr\'eal, H3C 3J7, Québec, Canada.}
\affiliation{Institut Trottier de Recherche sur les exoplan\`etes, Universit\'e de Montr\'eal}

\author{William Frost} 
\affiliation{D\'epartement de Physique and Observatoire du Mont-M\'egantic, Universit\'e de Montr\'eal, C.P. 6128, Succ. Centre-ville, Montr\'eal, H3C 3J7, Québec, Canada.}
\affiliation{Institut Trottier de Recherche sur les exoplan\`etes, Universit\'e de Montr\'eal}

\author[0000-0001-6758-7924]{Salma Salhi} 
\affiliation{D\'epartement de Physique and Observatoire du Mont-M\'egantic, Universit\'e de Montr\'eal, C.P. 6128, Succ. Centre-ville, Montr\'eal, H3C 3J7, Québec, Canada.}
\affiliation{Institut Trottier de Recherche sur les exoplan\`etes, Universit\'e de Montr\'eal}
\affiliation{Department of Physics and Astronomy, University of Calgary, 2500 University Dr NW, Calgary, AB T2N 1N4, Canada}

\author[0000-0002-2195-735X]{Louis-Philippe Coulombe} 
\affiliation{D\'epartement de Physique and Observatoire du Mont-M\'egantic, Universit\'e de Montr\'eal, C.P. 6128, Succ. Centre-ville, Montr\'eal, H3C 3J7, Québec, Canada.}
\affiliation{Institut Trottier de Recherche sur les exoplan\`etes, Universit\'e de Montr\'eal}

\author[0000-0001-5578-1498]{Bj\"orn Benneke} 
\affiliation{D\'epartement de Physique and Observatoire du Mont-M\'egantic, Universit\'e de Montr\'eal, C.P. 6128, Succ. Centre-ville, Montr\'eal, H3C 3J7, Québec, Canada.}
\affiliation{Institut Trottier de Recherche sur les exoplan\`etes, Universit\'e de Montr\'eal}

\author[0000-0003-4816-3469]{Ryan J. MacDonald}
\affiliation{Department of Astronomy, University of Michigan, 1085 S. University Ave., Ann Arbor, MI 48109, USA. NHFP Sagan Fellow}

\author[0000-0002-6773-459X]{Doug Johnstone}
\affiliation{NRC Herzberg Astronomy and Astrophysics, 5071 West Saanich Rd, Victoria, BC, V9E 2E7, Canada}
\affiliation{Department of Physics and Astronomy, University of Victoria, Victoria, BC, V8P 5C2, Canada}

\author[0000-0001-7836-1787]{Jake D. Turner}
\affiliation{ 
Department of Astronomy and Carl Sagan Institute, Cornell University, 122 Sciences Drive, Ithaca, NY 14853, USA
NHFP Sagan Fellow}

\author[0000-0002-5428-0453]{Marylou Fournier-Tondreau} 
\affiliation{D\'epartement de Physique and Observatoire du Mont-M\'egantic, Universit\'e de Montr\'eal, C.P. 6128, Succ. Centre-ville, Montr\'eal, H3C 3J7, Québec, Canada.}
\affiliation{Institut Trottier de Recherche sur les exoplan\`etes, Universit\'e de Montr\'eal}

\author[0000-0002-1199-9759]{Romain Allart}
\affiliation{D\'epartement de Physique and Observatoire du Mont-M\'egantic, Universit\'e de Montr\'eal, C.P. 6128, Succ. Centre-ville, Montr\'eal, H3C 3J7, Québec, Canada.}
\affiliation{Institut Trottier de Recherche sur les exoplan\`etes, Universit\'e de Montr\'eal}

\author[0000-0002-0436-1802]{Lisa Kaltenegger}
\affiliation{
Department of Astronomy and Carl Sagan Institute, Cornell University, 122 Sciences Drive, Ithaca, NY 14853, USA
}
